\documentclass[12pt]{iopart}
\usepackage{graphicx}
\usepackage{amssymb}
\usepackage{amscd}
\usepackage{iopams}
\usepackage[usenames]{color}
\usepackage{cite}
\usepackage{hyperref}

\newcommand{\nn}{\nonumber}

\newcommand{\rd}{{ \rm d }}
\newcommand{\J}{v}
\newcommand{\U}{c}
\newcommand{\g}{g}
\newcommand{\s}{s}
\def\tfrac12{\case{1}{2}}
\newcommand{\w}{\phi}
\newcommand{\uu}{\theta}
\newcommand{\R}{r}

\begin{document}
\jl{1}
\title[Bose-Hubbard dimers, Viviani's windows and pendulum dynamics]
{Bose-Hubbard dimers, Viviani's windows and pendulum dynamics }
\author{Eva-Maria Graefe$^{1}$,  Hans J\"urgen Korsch$^2$ and Martin P. Strzys$^2$}

\address{$^1$ Department of Mathematics, Imperial College London, London SW7 2AZ, UK }
\address{$^2$ FB Physik, TU Kaiserslautern, D--67653 Kaiserslautern, Germany}

\eads{\mailto{e.m.graefe@imperial.ac.uk}, \mailto{strzys@physik.uni-kl.de}, \mailto{korsch@physik.uni-kl.de}}
\begin{abstract}
The two-mode Bose-Hubbard model in the mean-field
approximation is revisited emphasizing a geometric interpretation
where the system orbits appear as intersection curves of
a (Bloch) sphere and a cylinder oriented parallel to the
mode axis, which provide a generalization of Viviani's curve
studied already in 1692.  In addition, the dynamics is  shown to agree with
the simple mathematical pendulum. The areas enclosed by the generalized Viviani curves, 
the action integrals, which can be used to semiclassically quantize the $N$-particle
eigenstates, are evaluated. Furthermore the significance
of the original Viviani curve for the quantum system is demonstrated. 
\end{abstract}

\submitto{\JPA}

\section{Introduction}
The two-mode Bose-Hubbard system is a popular
model in multi-particle quantum theory. It describes $N$ bosons, 
hopping between two sites with on-site interaction or a spinning 
particle with angular momentum $N/2$. Despite its simplicity, it offers
a plethora of phenomena and applications motivating an increasing number 
of investigations, somewhat similar to the harmonic oscillator in single-particle
quantum mechanics. In the limit of large $N$ it can be 
described by a non-linear Schr\"odinger or Gross-Pitaevskii equation
which generates a classical Hamiltonian dynamics 
(see, e.g.,~\cite{Milb97,Smer97,Ragh99,Vard01b,Angl01,Link06} 
for some studies related to the following).
Moreover it has been
shown recently \cite{07semiMP,Grae09dis,Bouk09,Chuc10,Niss10,Simo12}
that a semiclassical WKB-type construction can be
used to approximately recover quantum effects, such as the eigenstates 
and interference effects for finite (and even small) numbers of particles. This is based on
classical action integrals, as usual in semiclassics. 

In the present note, we point out an interesting relation between these
contemporary studies and much older ones by astronomers and mathematicians
from the times of Galilei or even earlier in ancient Greece as, e.g., 
Vincenzo Viviani (1622\,--\,1703) \cite{Viviani} or Eudoxus of Cnidus (408\,BC\,--\,355\,BC)
\cite{Hippo}.
The basic link between these studies is the family of closed curves on
the surface of a sphere and the surface area they enclose, appearing as phase space action 
integrals in the Bose-Hubbard dimer. 

The paper is organized as follows: In section \ref{s-viviani} we give a brief 
review of the background of Viviani curves, and introduce a generalization. 
We furthermore calculate the areas enclosed by the generalized Viviani curves. 
In section \ref{sec_BH_MF} we describe the Bose-Hubbard dimer and its 
mean-field approximation and show that the dynamical trajectories of the 
latter are given by generalized Viviani curves. We also show how the 
dynamics of the mean-field system can be directly related to that of a 
mathematical pendulum. Finally we analyze the relevance of the Viviani 
curve for the quantum Bose-Hubbard dimer. We conclude with a summary in 
section \ref{sec_sum}.

\section{Viviani curves}
\label{s-viviani}
In 1692 Vincenzo Viviani, a disciple of Galileo Galilei, proposed a problem
of the construction of four equal windows cut out of a hemispherical cupola
such that the remaining surface area can be exactly squared
\cite{Viviani} (see also \cite{cadd01} and references therein). The solution to this 
problem is given by an intersection of the hemisphere and a cylinder whose diameter
equals the radius of the hemisphere. This intersection curve of a sphere
and a cylinder tangent to the diameter of the sphere and its equator
is the famous Viviani curve, which can be generalized to an
intersection with an arbitrary cylinder.
For a recent extended study of such curves see \cite{Krol05,Krol07}.

\subsection{(Generalized) Viviani curves}
\label{ss-viviani}
\begin{figure}[t]
\begin{center}
\includegraphics[width=70mm]{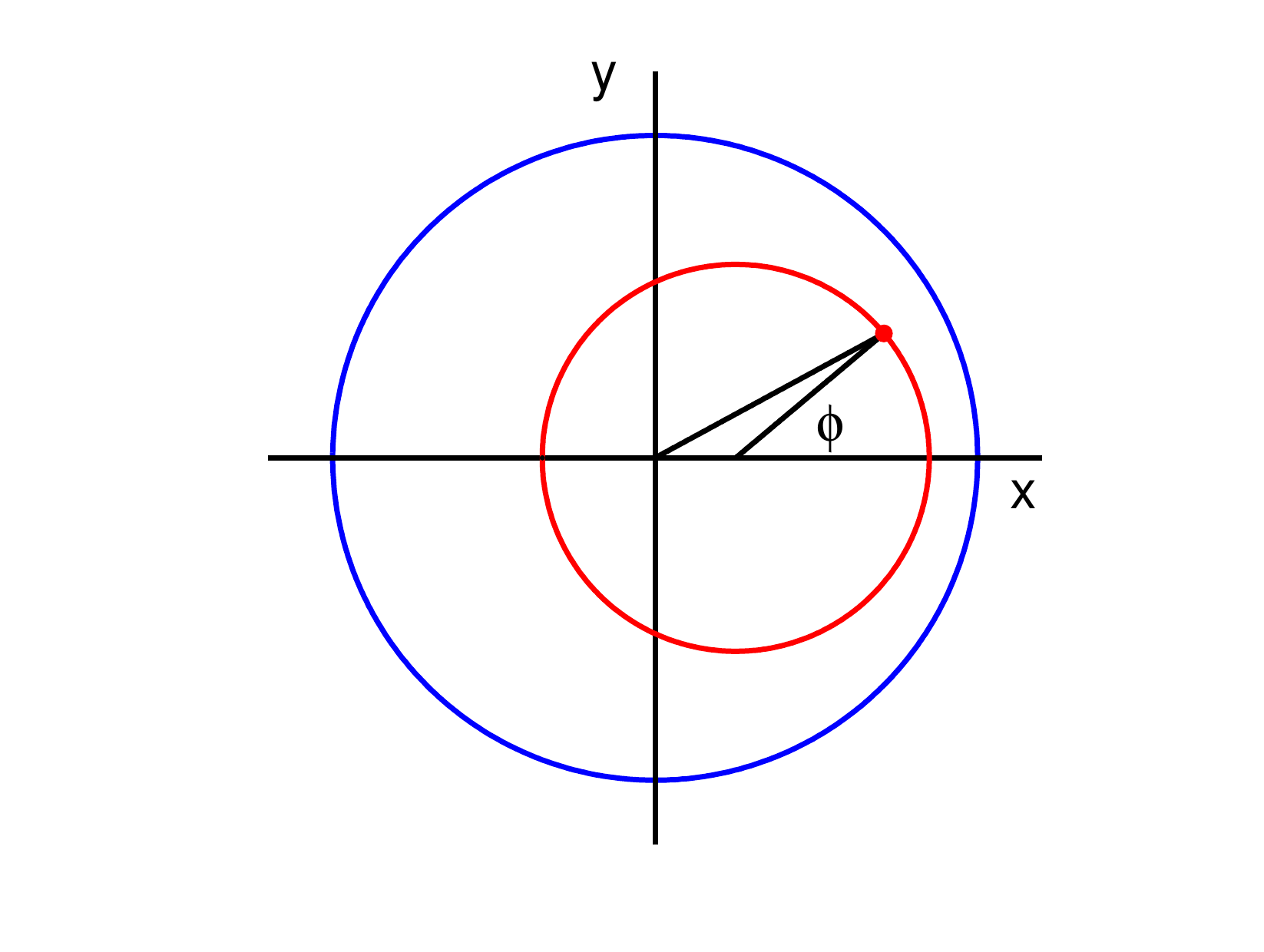}
\includegraphics[width=70mm]{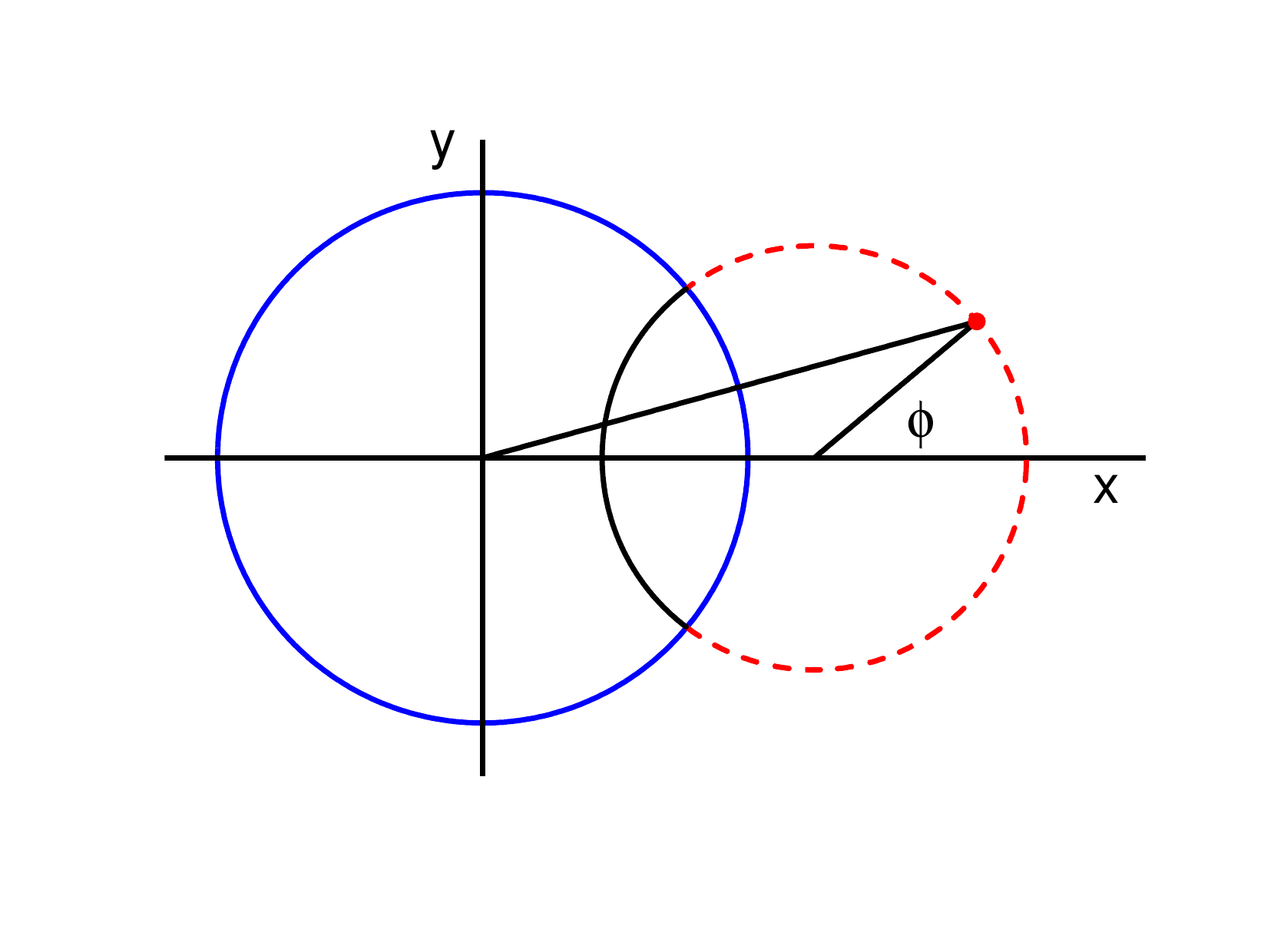}
\end{center}
\caption{\label{fig-circles12} Projection of the intersection of the unit sphere and the cylinder on the
$(x,y)$-plane for $a=0.25$, $\R=0.6$ (left) and  $a=1.25$, $\R=0.8$ 
(right). This corresponds to the dynamical trajectories of the mean-field Bose-Hubbard dynamics, where the dashed part of the circle is energetically inaccessible.} 
\end{figure}

Consider a cylinder of radius $\R$ centered around an axis in the $z$-direction 
shifted by $a$ in the $x$-direction:
\begin{equation}\label{eqn_cyl}
(x-a)^2+y^2 =\R^2.
\end{equation}
We will parametrize the basis of the cylinder 
by an angle $\w$,
\begin{equation}\label{eqn_circle}
x=a+\R\cos \w \ ,\quad y=\R\sin \w \,
\end{equation}
as shown in figure \ref{fig-circles12}.
\begin{figure}[b]
\begin{center}
\includegraphics[width=70mm]{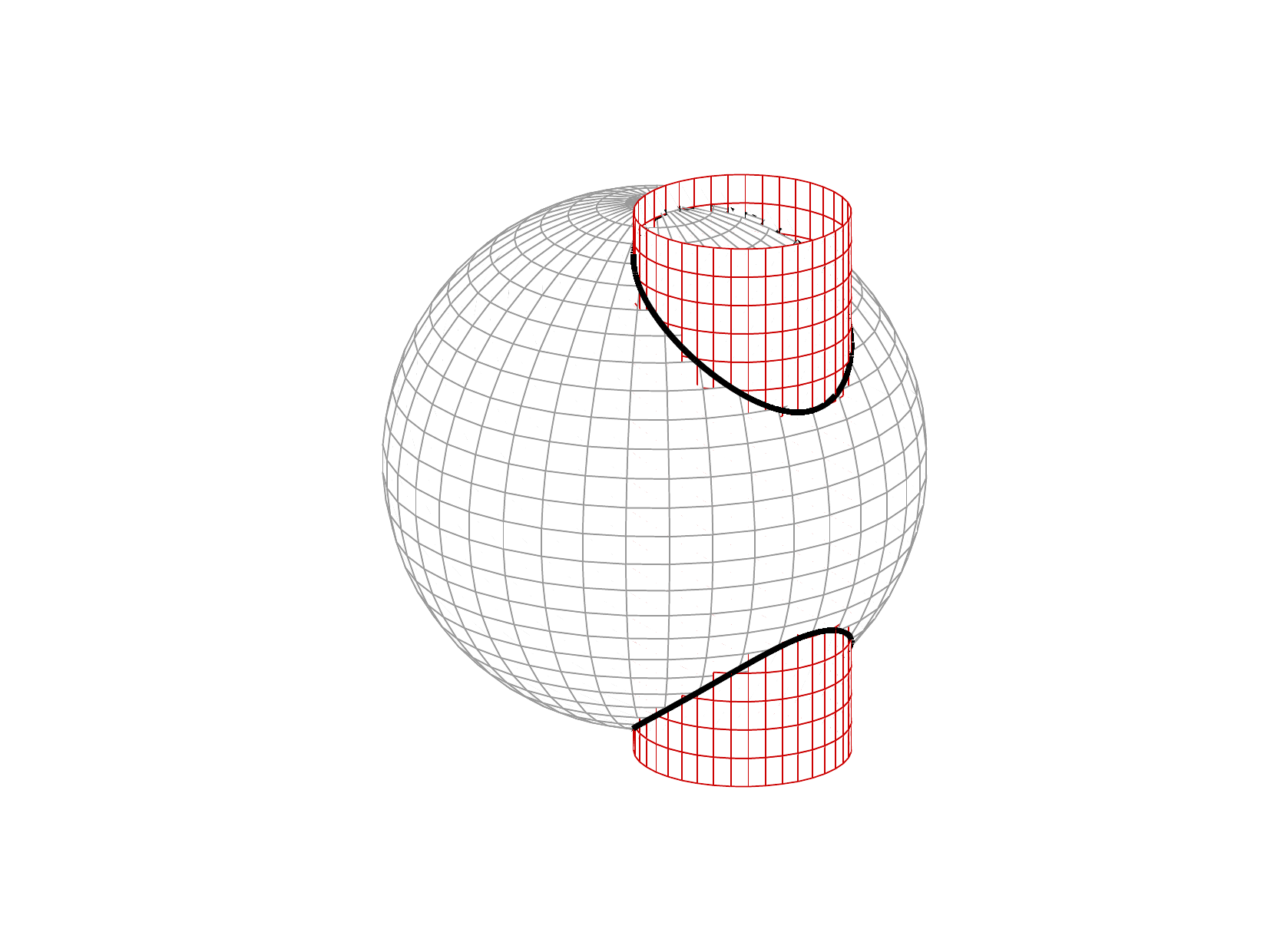}\hspace*{-18mm}
\includegraphics[width=70mm]{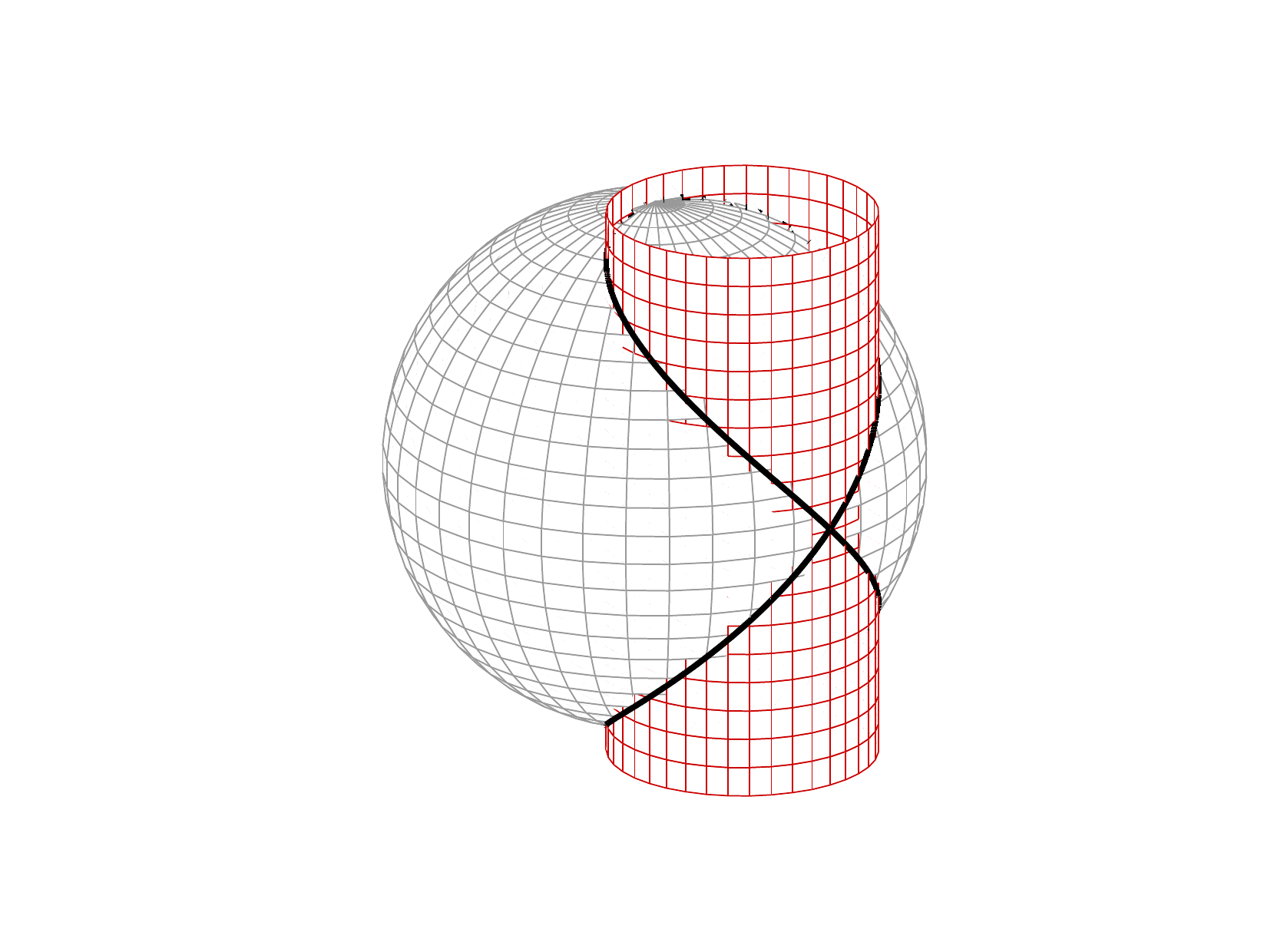}\hspace*{-18mm}
\includegraphics[width=70mm]{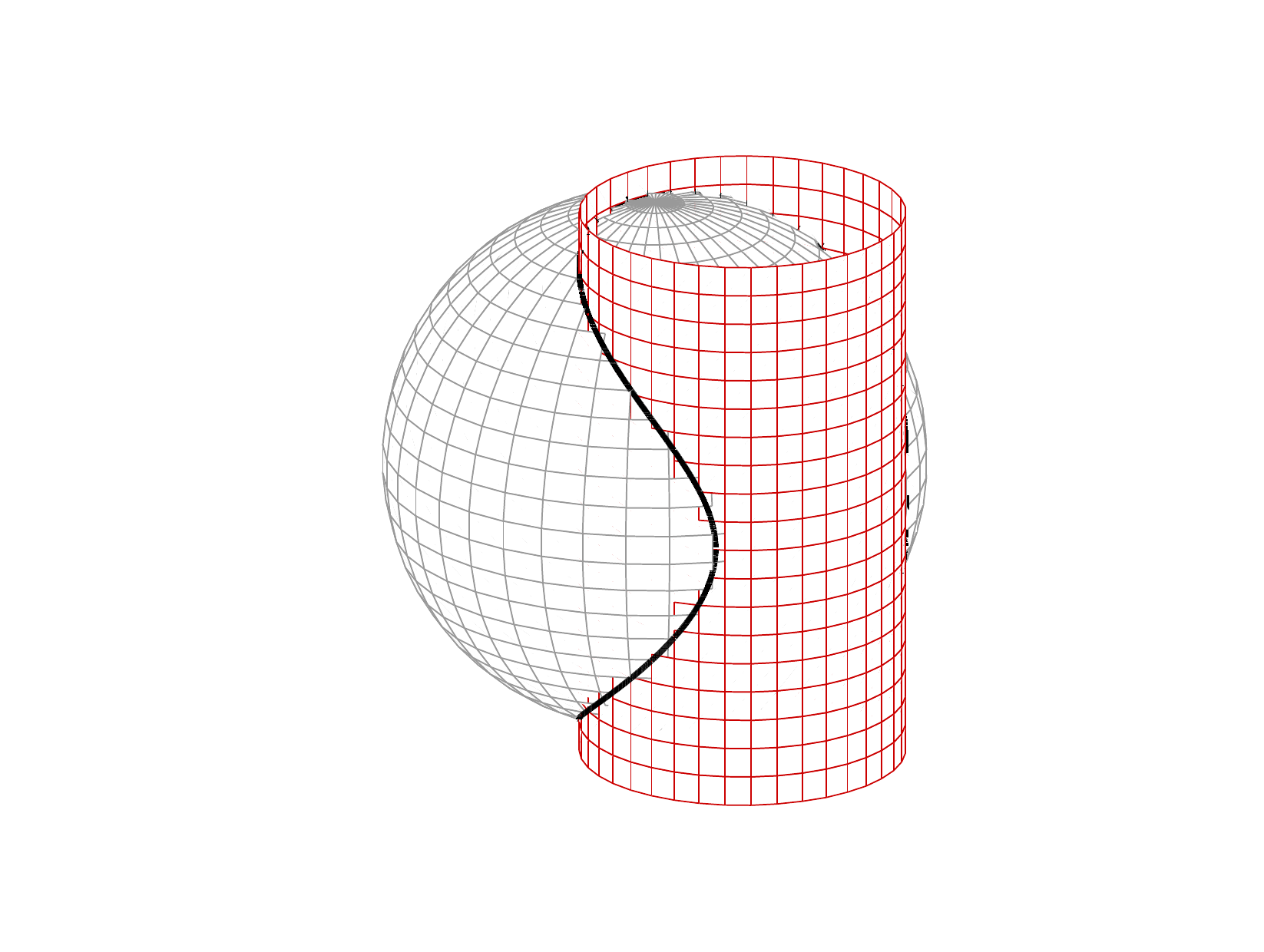}\\
\end{center}
\caption{\label{fig_viv_05} Intersection of the unit sphere and a cylinder displaced by
$a=0.5$ along the $x$-axis for radius $\R=0.4$, $0.5$ and $0.6$ (from left to
right). The figure in the middle is the Viviani curve.} 
\end{figure}

\begin{figure}[b]
\begin{center}
\includegraphics[width=80mm]{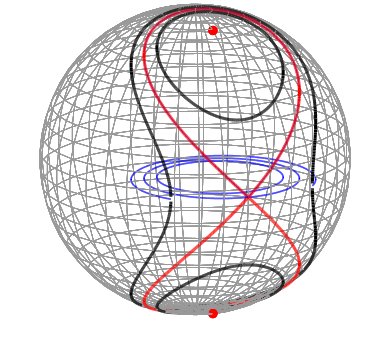}
\end{center}
\caption{\label{fig_vivianicurve} Viviani curve (red) and focal points (red
dots) together with two confocal euclidic spherical ellipses (black). Also show are
the projections on the $(x,y)$-plane.} 
\end{figure}

The generalized Viviani curves are given by the intersection curves of 
this cylinder with the unit sphere centered at the origin illustrated in figures
\ref{fig_viv_05} and \ref{fig_viv_07}. 
The radius $\R$ of the cylinder varies between $\R=0$ and
$\R=1+a$, otherwise there is no intersection with the sphere. 
For $\R=1+a$
the cylinder touches the sphere at $x=-1$.
We have to distinguish
two cases (see figure \ref{fig-circles12}):

\begin{figure}[t]
\begin{center}
\includegraphics[width=70mm]{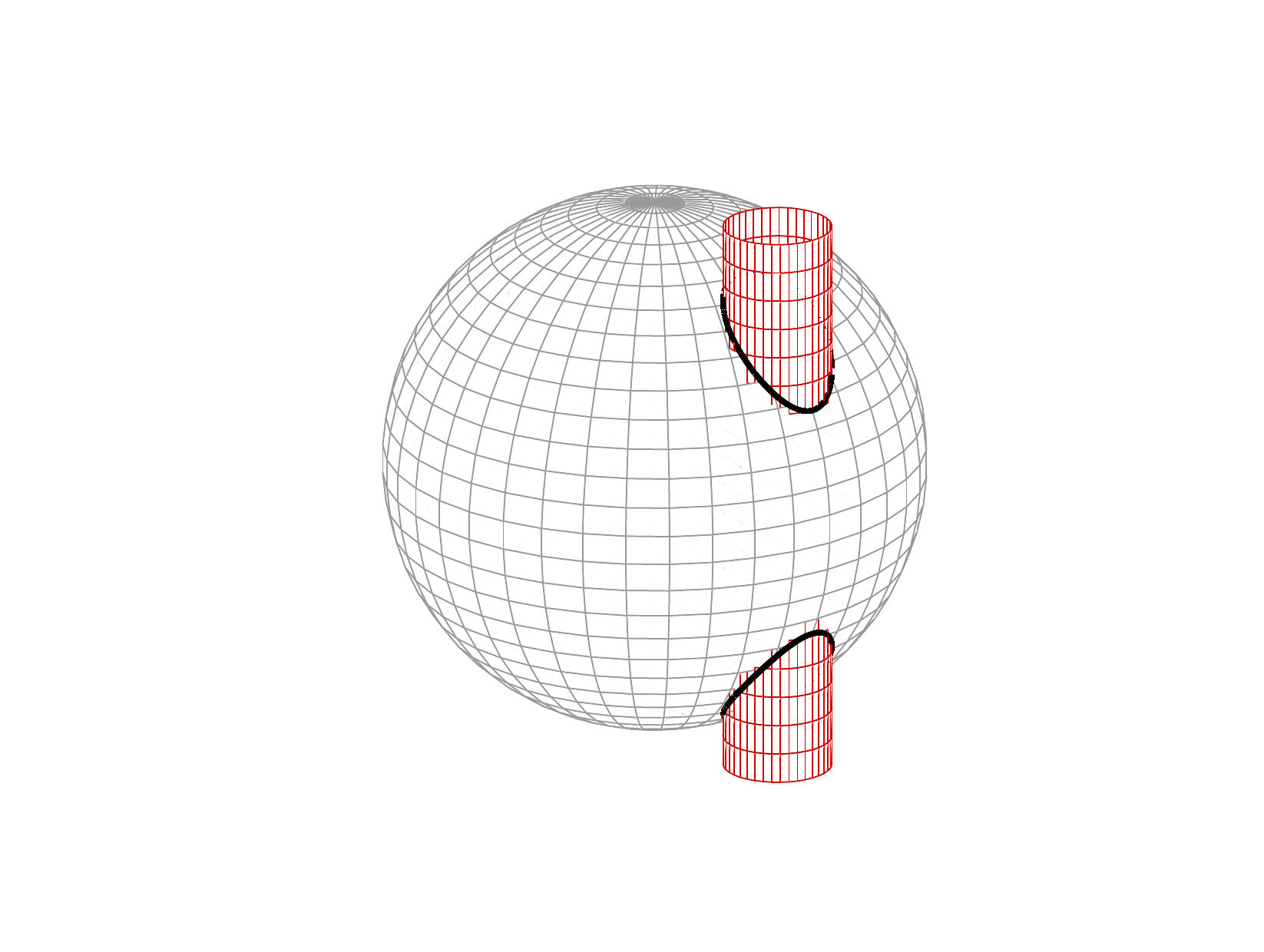}\hspace*{-18mm}
\includegraphics[width=70mm]{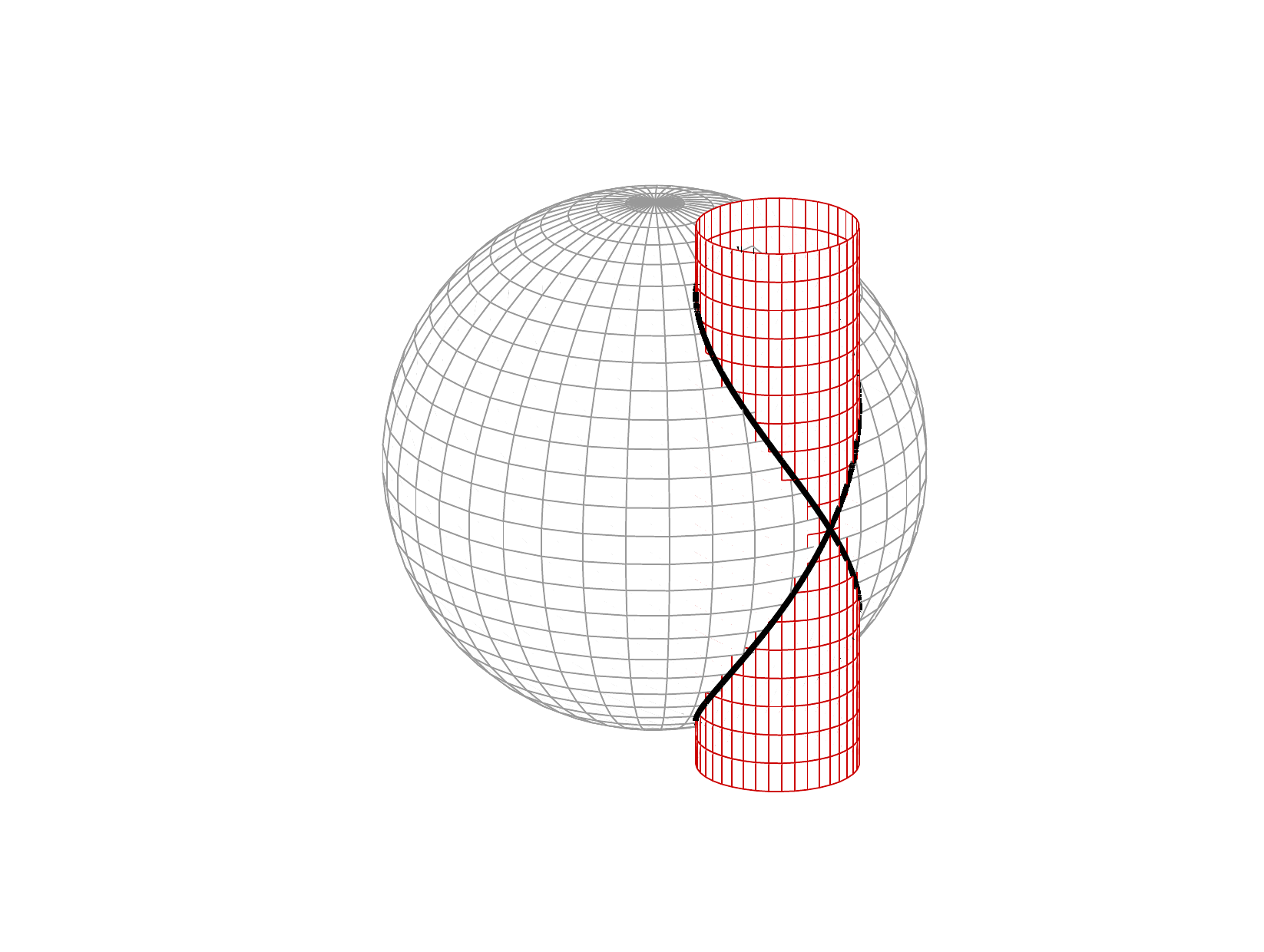}\hspace*{-18mm}
\includegraphics[width=70mm]{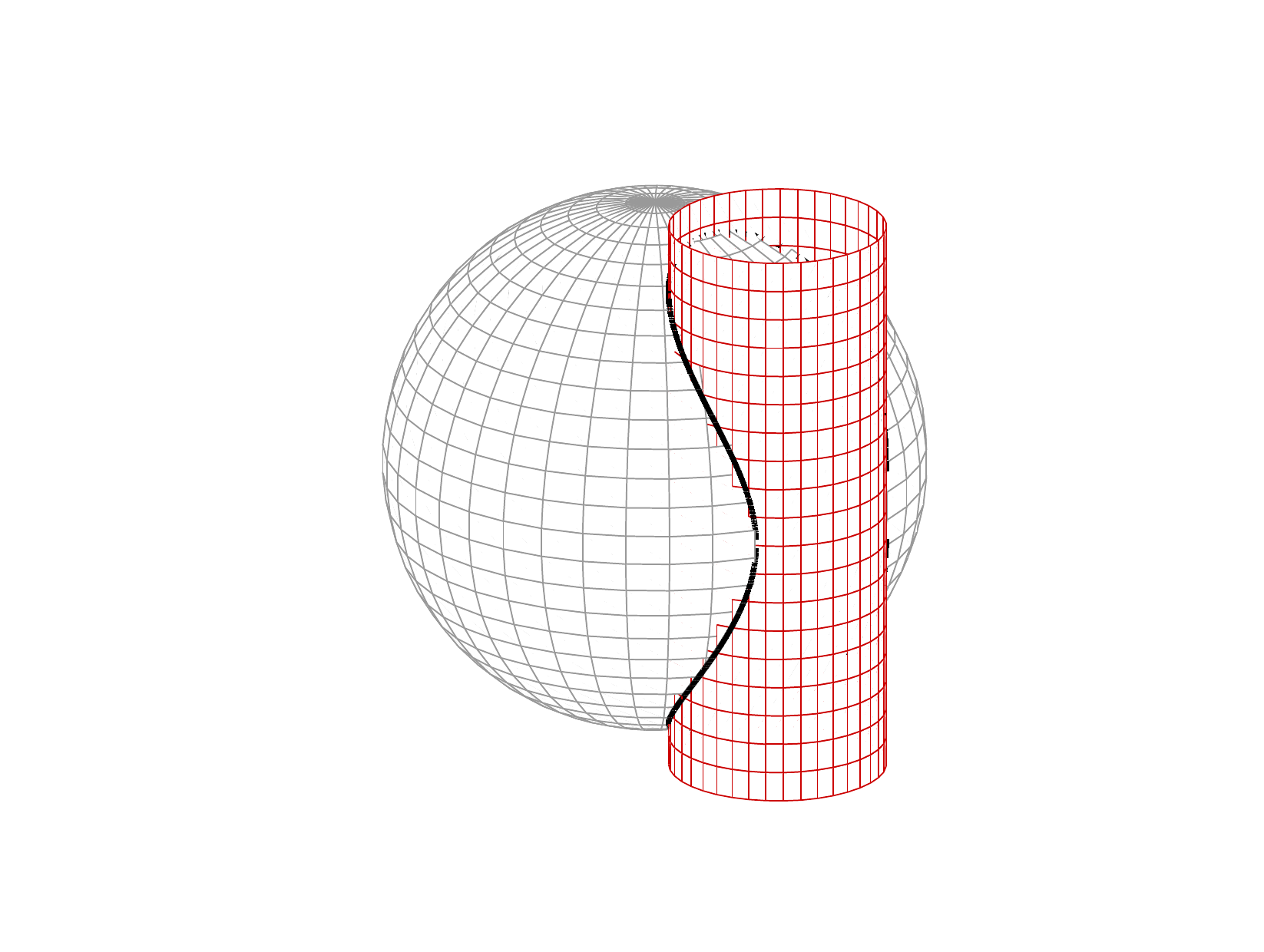}
\end{center}
\caption{\label{fig_viv_07} Intersection of the unit sphere and a cylinder displaced by
$a=0.7$ along the $x$-axis for radius $\R=0.2$, $0.3$ and $0.4$ (from left to
right).} 
\end{figure}
\begin{itemize}
\item[(i)] For $0< a<1$ and $0<\R <1-a$
all points on the circle (\ref{eqn_circle}) lead to
a full intersection of the cylinder and the sphere.
We obtain two single intersection loops on the
northern or southern hemisphere 
containing the points $(a,0,+\sqrt{1-a^2})$
and $(a,0,-\sqrt{1-a^2})$, respectively,
as illustrated on the left of figures \ref{fig_viv_05} and
\ref{fig_viv_07}.
\item[(ii)] For $0< a<1$ and $1-a<\R<1+a$ as well as for $a>1$ and $a-1<\R<1+a$
there is no intersection of the cylinder and the sphere
for $|\w |<\w_0$ with
\begin{equation}\label{w0}
\w_0=\arccos\frac{1-a^2-\R^2}{2a\R}\,.
\end{equation}
The curve of intersection with the sphere is a single loop
extending from the northern to the southern hemisphere (see figure \ref{fig_viv_05} and
\ref{fig_viv_07} (right)).
\end{itemize}
\begin{figure}[b]
\begin{center}
\includegraphics[width=55mm]{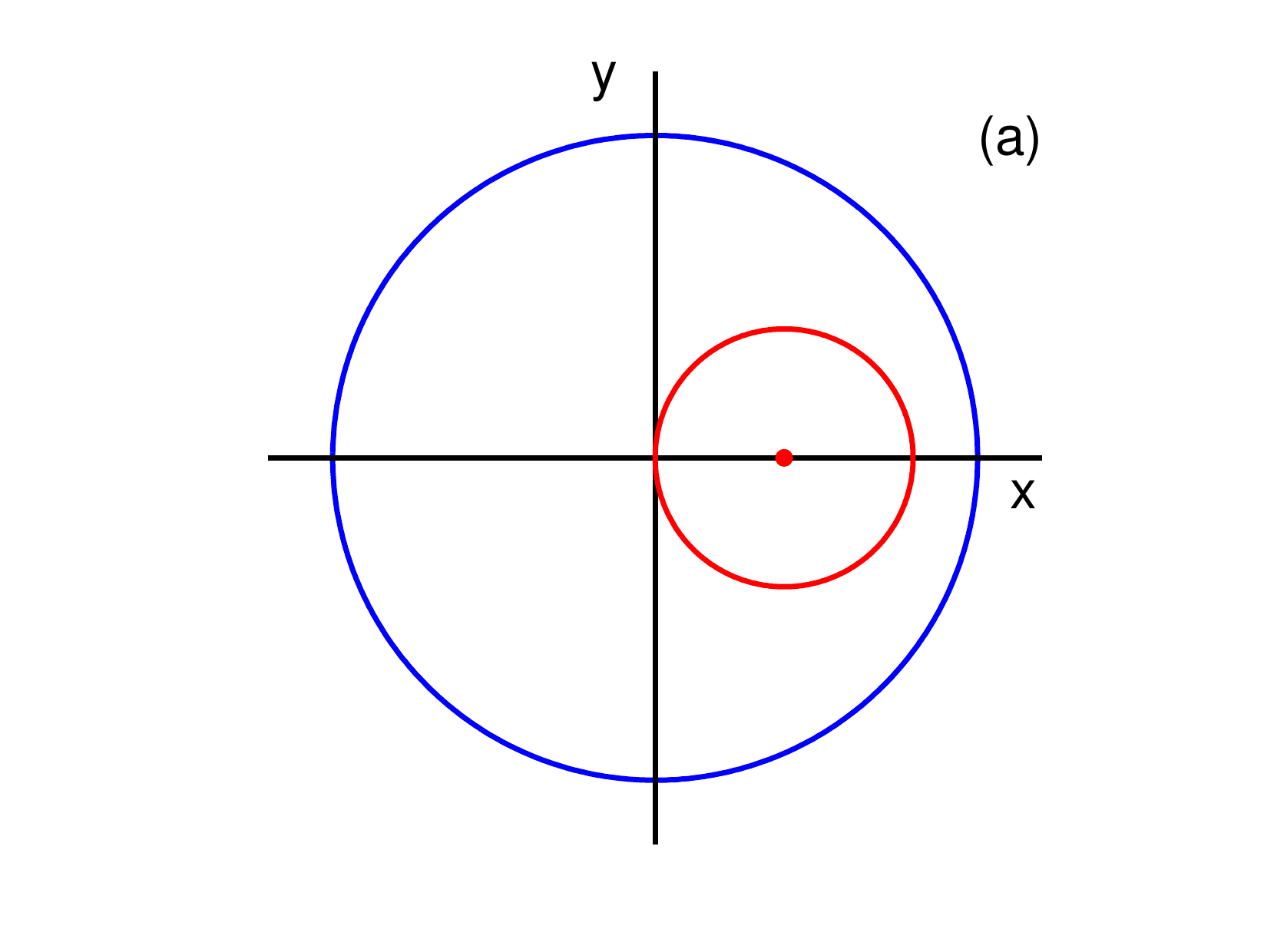}\hspace*{-8mm}
\includegraphics[width=55mm]{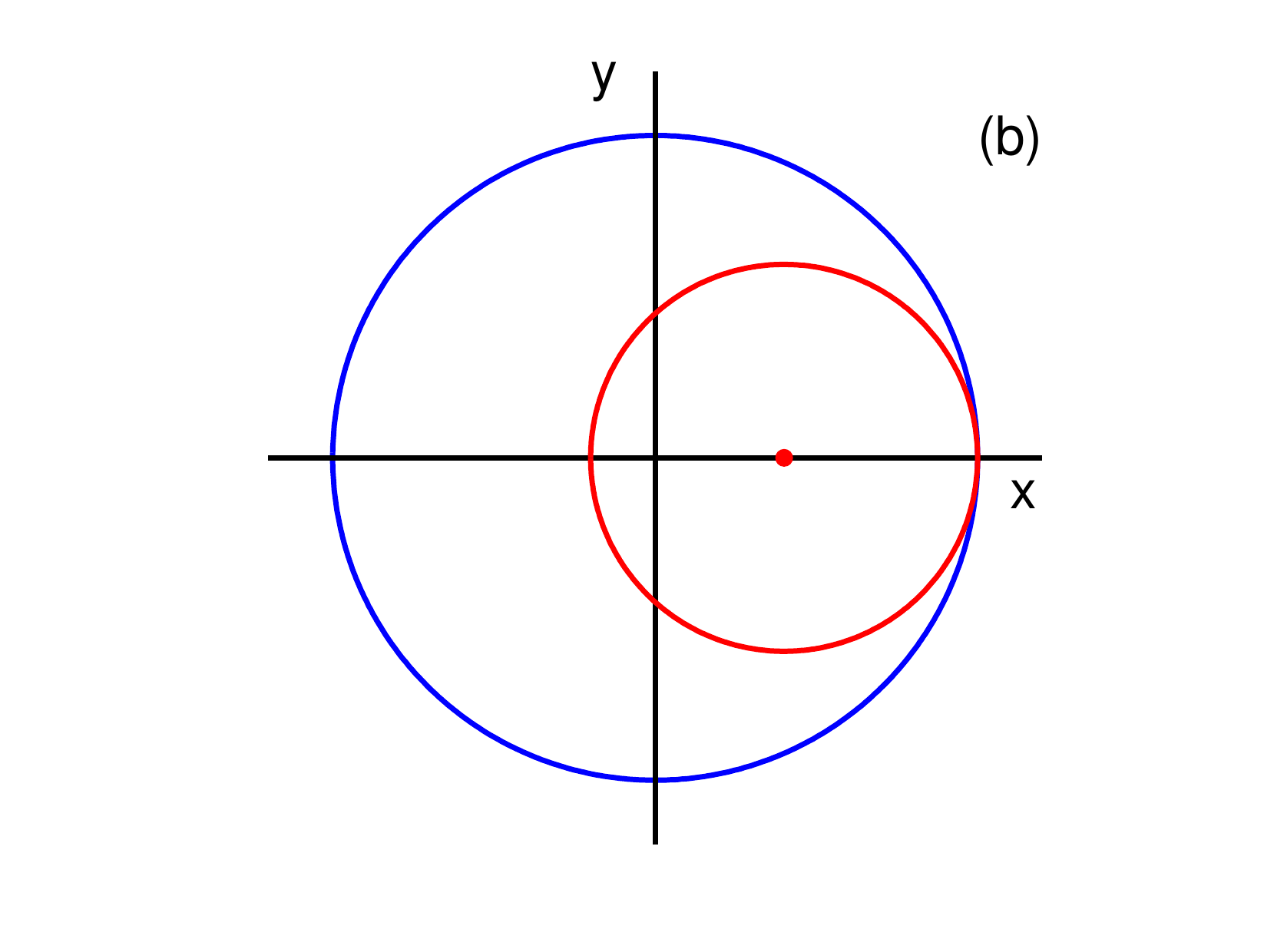}\hspace*{-8mm}
\includegraphics[width=55mm]{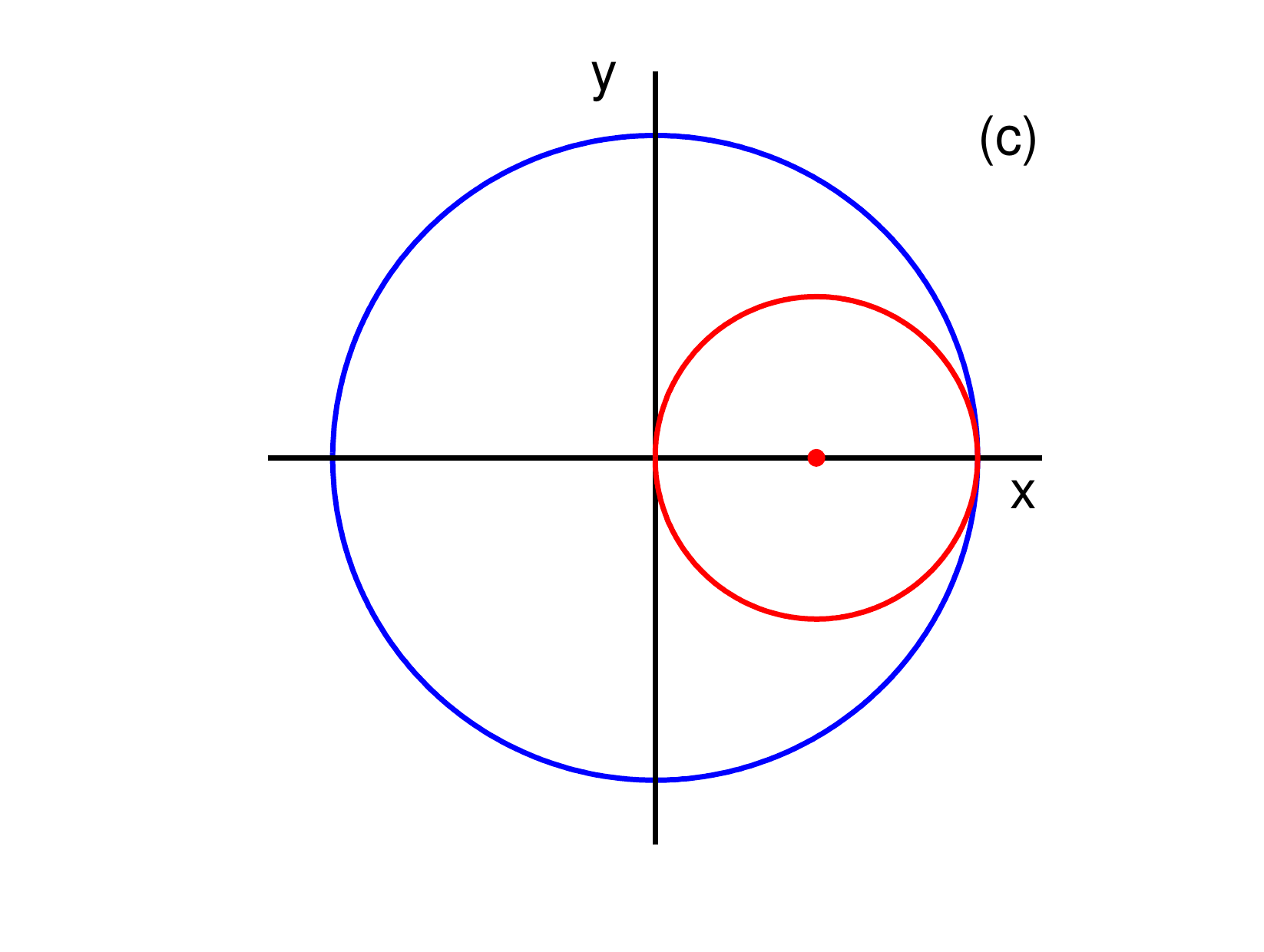}
\end{center}
\caption{\label{fig-circlesabc} Projection of the intersection of the sphere with a cylinder on the
$(x,y)$-plane for the special cases  $\R=a$ (case(a)). 
 $\R=1-a$ (case (b)) and  $\R=a=1/2$ (Viviani case (c)).} 
\end{figure}
\noindent
Three special situations are of interest (see figure \ref{fig-circlesabc}):
\begin{itemize}
\item[(a)] For $\R = a$ the cylinder passes through the center
of the sphere and the intersection loop passes through the
north- and south pole. As we shall see later, for the Bose-Hubbard dimer, this is the
frequently considered situation where the system is initially prepared in the
lower or upper state.
\item[(b)] For $\R =1-a$, which is only possible for $a\le 1$, the
cylinder is tangent to the sphere at the point ${\boldsymbol s}_{0+}=(1,0,0)$. 
This implies that the
intersection curve is a figure-eight loop with a self-intersection at
${\boldsymbol s}_{0+}$ (see figure \ref{fig_viv_07}  (middle)). This
curve is also known as the Hippopede of the Greek astronomer and mathematician
Eudoxus of Cnidus (408\,BC -- 355\,BC) \cite{Hippo,Yave01}.
In the Bose-Hubbard mean-field dynamics it appears as a separatrix curve, 
which separates the flow inside (for smaller values of $\R$)
from the flow outside (for larger values of $\R$), 
\item[(c)] In the most singular situation  cases (a) and (b) coincide,
i.e.~the cylinder passes through the center and touches the sphere.
This happens only for $\R=a=\tfrac12$ and is the original Viviani case
illustrated in figure \ref{fig_viv_05} (middle).
Using the parametrization (\ref{eqn_circle}) the Viviani curve
is given as
\begin{equation}\label{viviani-1}
x=\tfrac12+\tfrac12\cos \w\ ,\quad y=\tfrac12\sin \w
\end{equation}
and
$z^2=1-x^2-y^2= (1-\cos \w)/2=\cos^2(\w/2)$
and therefore 
$z=\pm \cos (\w/2)$\,.
This implies the relation \,$\varphi=\w/2$\, between the
azimuthal polar angle $\varphi$ and the angle $\w$,
which can also be found by purely geometric arguments. Therefore
the Viviani curve is given by
\begin{equation}\label{viviani-2}
(x,\,y,\,z)=(\cos^2\varphi,\,\cos \varphi \sin \varphi,\,\sin \varphi)\,.
\end{equation}
Comparing with spherical polar coordinates (see (\ref{eqn_polar}) below) we 
observe that
the Viviani curve (\ref{viviani-2}) can also be defined by
the simple condition that the azimuthal angle is equal to the polar angle measured from the equator:
\begin{equation}\label{viviani-3}
\varphi=\pi/2-\vartheta\,.
\end{equation}
\end{itemize}
In spite of the fact that only curve (\ref{viviani-2}) is the one originally
described by Viviani \cite{Viviani}, we will deliberately denote all the
intersection curves of a cylinder and a sphere as 
(generalized) {\it Viviani curves}.
Alternatively
these curves are known as {\it euclidic spherical ellipses}. This
is due to their remarkable property that the sum of the euclidic
distances to two focal points $x_F=a/(1+r)$, $y_F=0$, $z_F=\pm\sqrt{1-x_F^2}$
on the sphere equals a constant $2c$ with $c^2=z_F^2a/x_F$ \cite{Krol05}. These euclidean spherical ellipses
must be distinguished from the more popular version, where the distance
is measured by the arc length on the surface. The latter version found many applications in
navigation. Figure \ref{fig_vivianicurve} shows the original Viviani curve (\ref{viviani-2}),
whose focal points are at $(1/3,0,\pm \sqrt{8}/3)$, as well as two
generalized Viviani curves, i.e.~euclidic spherical ellipses, which
are chosen to possess the same focal points, however with
$c$ smaller or larger than the value $c=2/\sqrt{3}$ for the Viviani
curve. Let us point out again that these euclidic spherical ellipses
can appear as two disconnected loops.

Note also that the projection of the generalized
Viviani curves on the $(y,z)$-plane, the curves
\begin{equation}
(z^2-1+a^2-r^2)^2+4a^2y^2=4a^2r^2\,,
\label{yzprojection}
\end{equation}
are well-known functions, at least for $r=a$, where the cylinder
passes through the midpoint of the sphere (case (a)), they are
known as Cassini ovals and for $r=1-a$, where the cylinder
is tangent to the sphere, one obtains an eight-curve, also known
as Lemniscate of Gerono.

The previous considerations exclusively discussed the geometry of
the intersection curves. These curves can, however, be also generated 
by the trajectories of
a time evolution. For the (original) Viviani curve, such a dynamical
generation is often realized on the basis of eq.~(\ref{viviani-3})
by simply assuming a combined rotation along the circle of latitude,
$\vartheta=\pi/2-\omega t$, and the meridian, $\varphi=\omega t$.
Also the more general separatrix (b), the 
Hippopede of Eudoxus, was constructed in an effort to
explain the retrograde motion of the planets by a rotation of nested
spheres that share a common center with the same frequency 
around different axes \cite{Hippo}. As we shall see later, the same curves arise 
as the dynamical trajectories of the mean-field approximation for the Bose-Hubbard dimer. 

\subsection{Area integrals}
\label{s-area}
The area enclosed by a generalized Viviani curve, or more precisely 
the sum of the enclosed areas if this curve consists 
of two loops, is of particular interest. Historically, of course, because
this is the origin of the old pseudo-architectural problem posed and 
solved by Vincenzo Viviani in 1692, as stated already at the beginning of this section.
Today, in the context of
the Bose-Hubbard dimer discussed in section \ref{sec_BH_MF}, the area is the basis for 
a semiclassical quantization of the eigenvalues of the $N$-particle
Bose-Hubbard dimer \cite{07semiMP,Grae09dis,Simo12}. 

The calculation of an area $S$ enclosed 
by a curve on the sphere can be conveniently carried out by means of
the area conserving  projection of the unit sphere
onto a cylinder touching the sphere along the equator, which was already
known by Archimedes:
\begin{equation}
\label{cyl-pro}
(x,y,z) \longrightarrow (x',y',z')=(\cos \varphi, \sin \varphi, z),
\end{equation}
that is, 
\begin{equation}
\label{cyl-prob}
 \varphi=\arctan\, y/x\quad {\rm for}\quad z\ne \pm 1\,.
\end{equation}
The poles $(0,0,\pm 1)$ are projected onto circles $(\cos \varphi, \sin \varphi, \pm 1)$.
The area element is $\rd S=z(\varphi)\,\rd \varphi$,
where $z(\varphi)$ is the $z$-component of the points $(x,y,z)$ on the curve
parametrized by the azimuthal angle $\varphi$.

As a first example we calculate the area on the sphere enclosed by the Viviani curve.
The full area enclosed by the figure-eight-shaped Viviani curve is given by
\begin{equation}\label{S-viviani-10a}
S_V=2\pi-2S_0\,,
\end{equation}
where $S_0$ is the area between one half of the Viviani curve and the equator. 
From  the $z$-component given in (\ref{viviani-2}) we see that the cylinder
projection (\ref{cyl-pro}) maps the Viviani curve onto a sine-function and thus we find
\begin{equation}\label{S-viviani-1_0}
S_0=2\int_0^{\pi/2}\!\!z(\varphi)\,\rd \varphi=2\int_0^{\pi/2}\!\!\sin\varphi\,\rd \varphi=2\,.
\end{equation}
Note that on a more general sphere with radius $R$ this area is given by $4R^2$ 
as required from the solution to the original Viviani problem.
The area enclosed by the Viviani curve is thus given by
\begin{equation}\label{S-viviani-1}
S_V=2\pi-2S_0=2\pi-4\approx 2.2832\,.
\end{equation}

For evaluating the area integral in the general case, it is convenient to 
transform to the variable $\w $ by means of (\ref{eqn_circle}):
\begin{equation}
\fl \quad z(\varphi)\,\rd \varphi=z(\w)\,\frac{\rd \varphi}{\rd \w}\,\rd \w
=\R  \sqrt{1-\R^2-a^2-2a\R\cos\w \ }\ 
\frac{\R+a\cos \w}{a^2+\R^2+2a\R\cos\w }\,\rd \w,
\end{equation}
and the area outside the curve is given by
\begin{equation}\label{S0-1}
S_0=\R \int_{\w_0}^\pi \sqrt{1-\R^2-a^2-2a\R\cos\w }\ 
\frac{\R+a\cos \w}{a^2+\R^2+2a\R\cos\w }\,\rd \w\,,
\end{equation}
where the lower bound $\w_0$ is equal to zero for case (i) and
otherwise given by (\ref{w0}). The integrand is proportional to
$1/(r-a)$ for $\w=\pi$ so that we have an integrable singularity for
$\R=a$.
The area $S$ inside the curve is then equal to zero for
$a>1$, \ $\R<a-1$ and equal to $4\pi$ for $\R>1+a$, otherwise it is 
given by
\begin{equation}\label{S-gen}
S=\left\{
\begin{array}{ll}
-4S_0 \quad &{\rm for}\quad \R<a \\
2\pi -2S_0\quad &{\rm for}\quad   \R=a\\
4\pi -4S_0\quad &{\rm for}\quad   \R>a\, .
\end{array}\right.
\end{equation}
For the special cases distinguished above the area integral can be
evaluated in closed form:
\begin{itemize}
\item[(a)] For $\R=a$, the transformation from
$\varphi$ to $\w$ simplifies to $\varphi=\w/2$ and the integral
(\ref{S0-1}) to
\begin{eqnarray}
S_0&=& \int_{\w_0}^\pi \sqrt{1-2a^2-2a^2\cos\w \ }\,\rd \w \nn\\[2mm]
&=&\left\{\begin{array}{ll}
2E(4a^2) \ &{\rm for}\ a\le1/2\\[2mm]
4a\Big(E(1/4a^2)-\big(1-1/4a^2\big)K(1/4a^2)\Big)\ &{\rm for}\ a>1/2
\end{array}\right.\label{S0-3}
\end{eqnarray}
where $E(m)$ and $K(m)$ are complete elliptic integrals of the first
and second kind with parameter $m$.
\item[(b)]
For $\R=1-a$ (only possible for $a<1$, special case (b) mentioned above)
where the intersection curve is an eight-shaped curve with a double-point
at the fixed point ${\boldsymbol s}_{0+}=(1,0,0)$ where sphere and cylinder
touch each other, the area integral can be also calculated in
closed form with the amazingly simple result  \cite[sect.~3.2]{Krol07}
\begin{equation}\label{S-sep}
S=8\arcsin \sqrt{1-a\,}\ -8\sqrt{a(1-a)\,}\,.
\end{equation}
\item[(c)] For the Viviani case $\R=a=1/2$ both results 
agree with $S_V$ in (\ref{S-viviani-1}).
\end{itemize}

\begin{figure}[htb]
\begin{center}
\includegraphics[width=70mm]{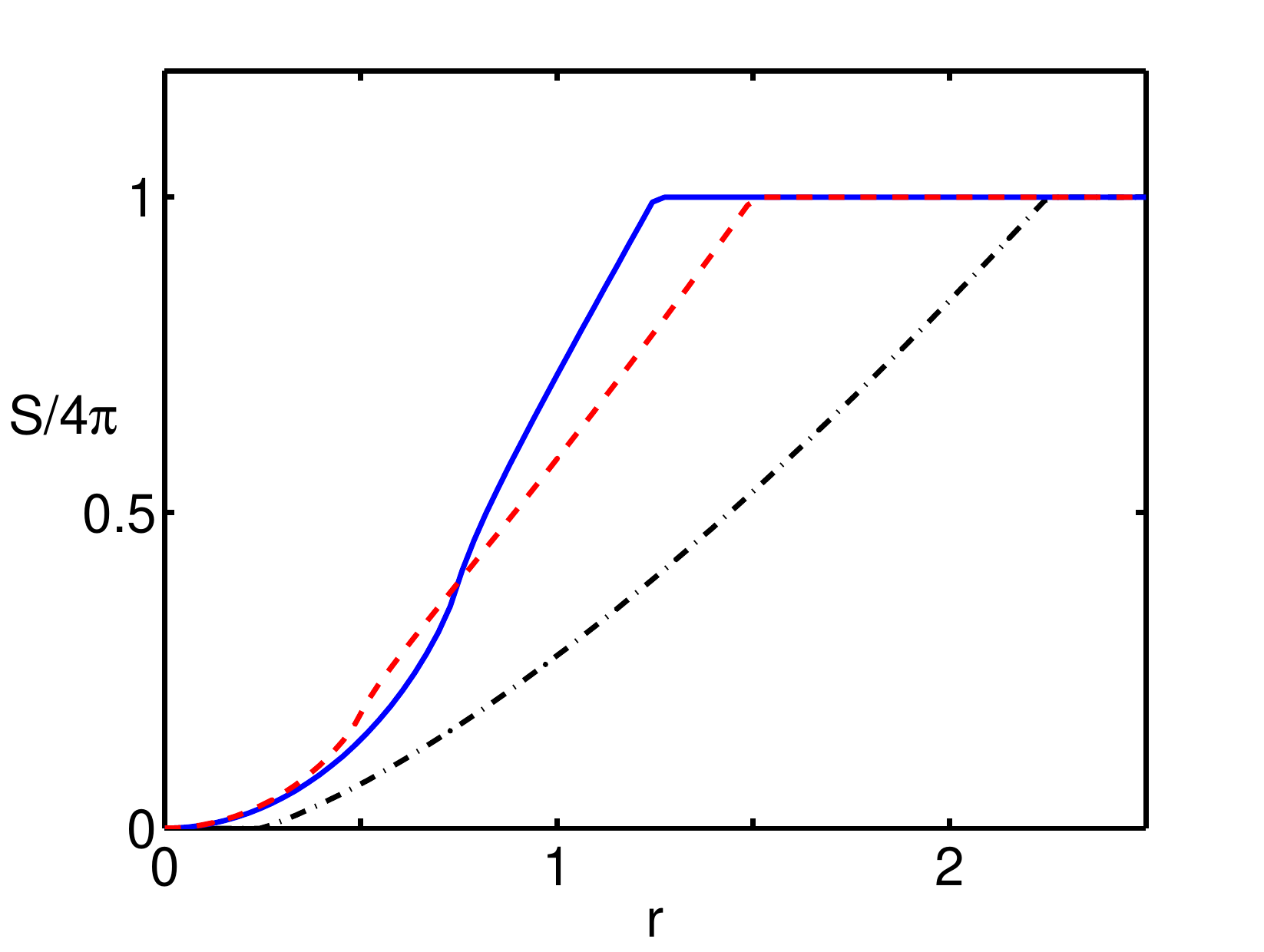}
\hspace*{10mm}
\includegraphics[width=70mm]{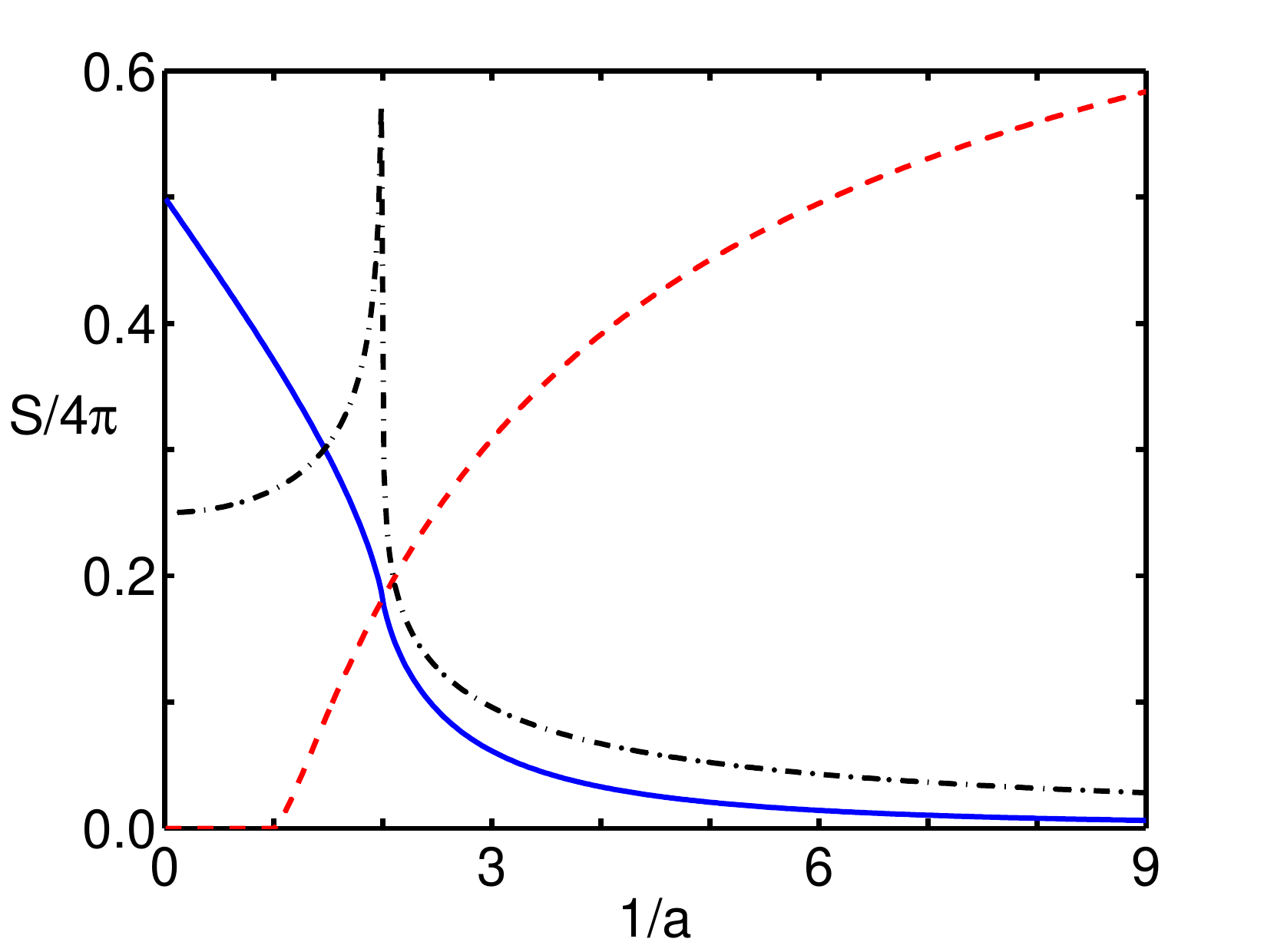}
\end{center}
\caption{\label{fig_S_rho_a} Left: Area $S/(4\pi)$ as a function of cylinder
radius $\R$ for $a=0.25$ (solid blue), $a=0.5$ (dashed red) and
 $a=1.5$ (dashed dotted black). 
 Right: Area $A=S/(4\pi)$ as a function of the
inverse cylinder displacement $a^{-1}$ for the critical
curve passing through the poles (solid blue) and for the critical figure-eight-shaped curve
passing through the point ${\boldsymbol s}_{0+}=(1,0,0)$ (dashed red), i.e.~eq.(\ref{S-sep}). 
The black dashed dotted curve shows the related time period $T/(4\pi)$ of the mean-field 
Bose-Hubbard dynamics for $v=1$ (see eq.~(\ref{periodphoa})).} 
\end{figure}
As an example, figure \ref{fig_S_rho_a} shows on the left the area $S$ 
enclosed by the
(generalized) Viviani loops as a function of the radius $\R$ of the cylinder 
for three values of the displacement $a$ chosen in the different regions.
In each case, the area increases monotonically from $0$ to $4\pi$.
The plot on the right shows the area as a function of the
inverse cylinder displacement $a^{-1}$ for the
curves (a) passing through the poles ($\R=a$). These pole trajectories
exist for all values of $a$. In the limit $a^{-1}\rightarrow 0$
the cylinder with $\R=a$ approaches the $(y,z)$-plane and intersects the
sphere in a great circle, which divides it 
into two equal hemispheres of area $2\pi$. With increasing $a^{-1}$ the 
intersection loop is deformed and the area enclosed shrinks, but 
it is still a single closed curve up to the critical point $a=1/2$, 
the Viviani case, where it bifurcates into two loops encircling the two
extrema.\\
Also shown in this figure is the area integral
for case (b), see eq.~(\ref{S-viviani-1}). These figure-eight shaped
trajectories only exist for $a<1$ and, 
with decreasing $a$, the area enclosed grows.
For $a=1/2$ the cases (a) and (b) coincide, and both curves pass through the
Viviani area $S/4\pi\approx 0.1817$ according to (\ref{S-viviani-1}).
For still smaller values of $a$ the cylinder center
approaches the center of the sphere and the area enclosed by the 
loops (a) through the poles goes to zero as $S\approx 2\pi a^2$ whereas the area enclosed by 
the loops through the touching point approaches the full surface $4\pi$.

We shall return to discussing the role of the area integrals for the quantum 
spectrum of the Bose-Hubbard dimer in section \ref{s-quantum}, after reviewing 
the quantum Hamiltonian and its mean-field approximation in the following.
\section{The Bose-Hubbard dimer and the mean-field approximation}
\label{sec_BH_MF}
The two-mode Bose-Hubbard system, describing cold bosonic atoms 
on a `lattice' consisting of only two sites, is a standard model in the field of cold atoms
\cite{Milb97,Smer97,Ragh99}. It is described by the Hamiltonian 
\begin{equation}
\hat H=\epsilon(\hat n_1  -\hat n_2)
 + \J (\hat a_1^\dagger \hat a_2 + \hat a_1 \hat a_2^\dagger)
 +\tfrac12 \U (\hat n_1  - \hat n_2)^2,
 \label{BH-Hamiltonian}
\end{equation}
with mode energies  $\pm\epsilon$, coupling  $v$ and
on-site interaction $\U$. The $\hat a_j$ and $\hat a_j^\dagger$ denote particle annihilation and creation operators in mode $j$ respectively. The total particle number $\hat N=\hat n_1+\hat n_2=\hat a_1^\dagger\hat a_1+\hat a_2^\dagger\hat a_2$ is a conserved quantity. Despite its simple theoretical structure the Bose-Hubbard dimer is of considerable
importance also from an experimental point of view, describing for example ultracold
atoms in a double-well trap or in the ground state of an external trap with two internal degrees of freedom \cite{Albi05,Zibo10}.

Introducing self-adjoint angular momentum operators $\hat L_x$, $\hat L_y$ and
$\hat L_z$ according to the Schwinger representation 
\begin{eqnarray}
\nonumber \hat L_x =\tfrac12(\hat a_1^\dagger\hat a_2+\hat a_1\hat a_2^\dagger)\,,\\ 
\label{lxlylz}\hat L_y=\case{1}{2\rmi}\,(\hat a_1^\dagger\hat a_2-\hat a_1\hat a_2^\dagger)\,,\\
\nonumber\hat L_z=\tfrac12(\hat
a_1^\dagger\hat a_1-\hat a_2^\dagger\hat a_2), 
\end{eqnarray}
with  $SU(2)$ commutation relation $[\hat{L}_x,\hat{L}_y]={\rm i}\hat{L}_z$
and cyclic permutations, the Hamiltonian
(\ref{BH-Hamiltonian}) can be written as:
\begin{equation}\label{BH-hamiltonian-L}
\hat H=2 \epsilon\hat L_z+2\J \hat L_x+2\U \hat
L^2_z.
\end{equation}
Here the conservation of the particle number $N$
appears as the conservation of
$L^2=\frac{N}{2}\big(\frac{N}{2}+1\big)$, i.e. the rotational
quantum number $L =N/2$. Note that the Hamiltonian (\ref{BH-hamiltonian-L}) is also
known as the Meshkov-Lipkin-Glick Hamiltonian, an angular momentum model originally
introduced in the context of nuclear physics as a solvable model against which 
to check typical approximation schemes of many-particle physics \cite{Lipk65,Mesh65,Glic65}. 
The mean-field approximation 
in the context of cold atoms is closely related to the classical approximation for 
the Lipkin-Meshkov-Glick system, which has been the subject of many studies 
in applications to quantum information theory  or quantum phase transitions
in the past (see, e.g., \cite{Vida04b,Lato05,Ribe07,Orus08} and references therein).

Both the static and dynamic properties of the full many-particle system (\ref{BH-Hamiltonian}) 
can be numerically deduced in a straight forward manner. This makes it an ideal model for the investigation of the correspondence to the approximate mean-field description, which we shall briefly discuss in the following.

\subsection{Mean-field dynamics}
The mean-field approximation can be formally obtained by replacing 
the bosonic operators by
c-numbers: $\hat a_j\to \psi_j$, $\hat a_j^\dagger\to \psi_j^*$.
Taking the macroscopic limit $N\to\infty$ with 
$cN=g=const.$ yields the discrete nonlinear Schr\"odinger or
Gross-Pitaevskii equation
\begin{eqnarray}
\rmi\dot{\psi_1}&=&\left(\epsilon+g(|\psi_1|^2-|\psi_2|^2)\right)\psi_1+v\psi_2\nn\\
\label{eqn-GPE-herm}
\rmi \dot{\psi_2}&=& v\psi_1-\left(\epsilon+g(|\psi_1|^2-|\psi_2|^2)\right)\psi_2,
\end{eqnarray}
for $\hbar=1$, where $\g=\U N$ is the overall interaction
between the particles.

Similar to linear time-dependent Schr\"odinger equations also the nonlinear dynamics (\ref{eqn-GPE-herm}) possesses a canonical structure \cite{Dira27,Wein89}: It can be derived from a
classical Hamiltonian function 
\begin{eqnarray}
H_{\rm cl}=\epsilon(|\psi_1|^2-|\psi_2|^2)+v(\psi_1^*\psi_2+\psi_1
\psi_2^*)+\frac{g}{2}(|\psi_1|^2-|\psi_2|^2)^2,
\label{classical-E-psi}
\end{eqnarray}
with canonical equations $\rmi\dot \psi_j=\partial H_{\rm cl}/\partial\psi_j^*$.

One can also formulate the nonlinear two-mode system in terms of the Bloch
vector of a spin-$\tfrac12$ system, introducing the spin components 
\begin{eqnarray}
\nonumber s_x=\tfrac12 \big(\psi_1^*\psi_2+\psi_1\psi_2^*\big)\,,\\ 
\label{sxsysz} s_y= \case{1}{2\rmi} \big(\psi_1^*\psi_2-\psi_1\psi_2^*\big)\,,\\
\nonumber s_z=\tfrac12 \big(\psi_1\psi_1^*-\psi_2\psi_2^*\big)\,,\ 
\end{eqnarray}
in accordance with
(\ref{lxlylz}). In these variables the total energy
takes the form
\begin{equation}\label{eqn_Hamfct-s}
H_{\rm cl}=2\epsilon\s_z+2v\s_x+2\g s_z^2,
\end{equation}
and the nonlinear Schr\"odinger
equation (\ref{eqn-GPE-herm}) translates to nonlinear Bloch equations of the form
\begin{eqnarray}
\dot \s_x &=& -2\epsilon \s_y-4\g \s_y\s_z\nn\\
\dot \s_y &=& 2\epsilon \s_x+4\g\s_x\s_z -2v \s_z \\
\dot \s_z &=& 2v \s_y\nn.
\label{bloch-eq-s}
\end{eqnarray}
These equations conserve the normalization, that is, the motion of the spin 
vector ${\boldsymbol s}=(\s_x,\s_y,\s_z)$ is confined to the surface of the 
Bloch sphere with radius $|{\boldsymbol s}|=1/2$.

Alternatively one can investigate the system in terms of the
coordinates $p=|\psi_1|^2-|\psi_2|^2$ and $q=(\arg{(\psi_2)}-\arg{(\psi_1)})/2$, which are related to 
the coordinates on the Bloch sphere via
\begin{eqnarray}\label{pq-transf}
p=2\s_z \ ,\quad 2q=\arctan \s_y/\s_x\,. 
\end{eqnarray}
That is, they are similar to the (area preserving) projection of the vector ${\boldsymbol s}$ on a cylinder touching the sphere at the equator as introduced before in equations (\ref{cyl-pro}) and (\ref{cyl-prob}).

The classical energy, that
is, the Hamiltonian function, in terms of $p$ and $q$ reads
\begin{eqnarray}
H_{\rm cl}=\epsilon p+v\sqrt{1-p^2}\cos(2q)+\frac{g}{2}p^2,
\label{classical-E-pq}
\end{eqnarray}
and the equations of motion (\ref{eqn-GPE-herm})
\begin{eqnarray}
\label{eq-motion-p}
\dot{p}&=&2v\sqrt{1-p^2}\sin(2q)\\
\label{eq-motion-q}
\dot{q}&=& \epsilon-v\frac{p}{\sqrt{1-p^2}}\cos(2q)+gp,
\end{eqnarray}
are again canonical, that is, $\dot q=\partial H_{\rm cl}/\partial p$,
$\dot p=-\partial H_{\rm cl}/\partial q$. This describes the motion of a 
pendulum whose length, however, depends dynamically on the momentum $p$. 
The energy $E=H_{\rm cl}$ is conserved. 

In the following we will confine ourselves to
the case of a symmetric Bose-Hubbard dimer, $\epsilon=0$, which captures important characteristics of the 
dynamics. In addition we will assume $v>0$ without loss of generality.
In this case, the fixed points of the dynamics of the Bloch vector (\ref{bloch-eq-s}), corresponding to stationary solutions of the Gross-Pitaevskii equation,
are given by
\begin{equation}
s_y=0\ ,\quad
s_z\in\Big\{0,\, 0,\, \pm\tfrac12\sqrt{1-v^2/g^2}\, \Big\}.
\end{equation}
Thus we have two fixed points for $|g|\leq v$ with
energies $E=-v$ and  $E=v$ which are a minimum and a maximum of the
energy surface. If the interaction strength $|g|$ is increased one
of these extrema bifurcates at the critical interaction $|g|=v$
into a saddle point, still at the equator, and two
extrema located away from the equator at $s_z=\pm\tfrac12\sqrt{1-v^2/g^2}$ with energy $E_{\rm m}=\frac{g}{2}(1+v^2/g^2)$, two maxima
for repulsive interaction $g>0$ or two minima for attractive
interaction. In both cases the other extremum (minimum 
with energy $E=-v$ for $g>0$ or maximum with energy $E=v$
for $g<0$) remains at the equator. In this supercitical case
the two fixed points $\s_z\ne 0$ correspond to stationary states mainly 
populating one of the levels, the \textit{self trapping} states.

For the following discussion it will be convenient to
rescale the spin components as $x=2\s_x$,  $y=2\s_y$, $z=2\s_z$
where the nonlinear Bloch equations (\ref{bloch-eq-s}) 
 in the symmetric case,
\begin{eqnarray}
\dot x &=& -2\g yz\nn\\
\dot y &=& +2\g x z -2v z \label{bloch-eq-r}\\
\dot z &=& 2v y\,,\nn
\end{eqnarray}
restrict the motion of the vector ${\boldsymbol s}=(x,y,z)$
to the unit sphere $|{\boldsymbol s}|^2=x^2+y^2+z^2=1$. 
We will assume that the parameters $v$ and $g$ are non-negative,
otherwise the dynamics can be obtained by simple symmetry arguments. We will also use
spherical polar coordinates 
\begin{equation}\label{eqn_polar}
(x,y,z)=(\sin \vartheta \cos \varphi,\sin \vartheta \sin \varphi,\cos
\vartheta)\,.
\end{equation}
Of basic importance for the structure of the flow on the sphere are the fixed points
\begin{equation}\label{fixpoint}
{\boldsymbol s}_{0\pm}=(\pm 1,0,0) \quad {\rm and}   \quad {\boldsymbol s}_{1\pm}=(a,0,\pm\sqrt{1-a^2}) 
\quad {\rm for} \quad a^2<1
\end{equation}
with $a=v/g$. The energy
\begin{equation}\label{eqn_energy}
E=vx+\tfrac12\,\g\,z^2 
\end{equation}
(see eq.~(\ref{eqn_Hamfct-s})) is conserved and the fixed points ${\boldsymbol s}_{0\pm}$
correspond to a maximum and a minimum energy $E_{0\pm}=\pm v$. At $a=1$
the maximum bifurcates into a saddle point at ${\boldsymbol s}_{0+}$
and two maxima at ${\boldsymbol s}_{1\pm}$ with energy $E_{1\pm}=g(1+a^2)/2$ in the self trapping transition.

It is now easy to see that the dynamical trajectories (\ref{bloch-eq-r}) are exactly given by 
the generalized Viviani curves introduced in section \ref{ss-viviani}:
Inserting $z^2=1-x^2-y^2$ we can rewrite equation (\ref{eqn_energy}) in 
the form (\ref{eqn_cyl}) with 
\begin{equation}\label{eqn_a_rho}
\R=\sqrt{1+a^2-2E/g\,}\,.
\end{equation}
Thus, the trajectories generated by the flow (\ref{bloch-eq-r})
can be interpreted as curves of intersection of the unit
sphere with a cylinder as illustrated in figures
\ref{fig_viv_05} and \ref{fig_viv_07}, that is, Viviani curves.

Of special interest for the two-mode
Bose-Hubbard dynamics is the case where initially the system is in the
lower or upper state, i.e., at one of the poles on the Bloch sphere.
This imposes the condition $\R=a$. For $a>\tfrac12$ (i.e. $g<v/2$) 
this trajectory traces out a closed euclidic ellipse on the
sphere. For $a=\infty$ this is a great circle in the $(y,z)$-plane
which tightens if $a$ is reduced. Note that the intersection points of this
ellipse with the equator at $x=1/2a$, $y=\pm\sqrt{1-1/(4a^2)\,}$ 
approach each other until they meet for $a=\tfrac12$ at the
fixed point ${\boldsymbol s}_+=(1,0,0)$, a double point of the 
figure eight shaped trajectory. 
For $a<\tfrac12$ the ellipse consists of two loops on the northern and
southern hemisphere which end up as circles around the poles for $a=0$. 

Let us now return to the time dependence generated
by the flow (\ref{bloch-eq-r}). The motion is periodic with a 
period $T$ derived by several authors before (see, e.g.,
\cite{Kenk86,Holt01a,Simo12}). Here we present a very simple derivation,
which also provides new insight into the dynamics.

For the simplest case, $g=0$, the flow equations (\ref{bloch-eq-r}) are linear. 
The $x$-component is conserved, $x=x_0$,
and we find a global rotation around the $x$-axis with frequency $\omega_{0x}=2v$.
In our picture this corresponds to the limit $a,\R\rightarrow \infty$ with fixed
$a-\R=x_0$.
Somewhat more interesting is the case $v=0$, Here the $z$-component is conserved, $z=z_0$, 
and we find a rotation around the $z$-axis. Here, however, 
we have
\begin{equation}
\dot x=-2gz_0y\ , \quad \dot y=2gz_0\,x\quad   \Rightarrow \quad \ddot x=-2gz_0\,\dot y =-4g^2z_0^2\,x 
\end{equation}
and the frequency
\begin{equation}
\omega_{0z}= 2g|z_0|=2g\sqrt{1-r^2}
\end{equation}
varies from a maximum  $\omega=2g$  at the poles to $\omega=0$ at the equator.

For the general case, inserting the time derivative of
transformation (\ref{eqn_circle}) into the flow equations (\ref{bloch-eq-r})
we find
\begin{equation}\label{dgl-penulum}
\dot \w=2gz=\pm 2g\sqrt{1-a^2-\R^2-2a\R \cos \w\,}\,,
\end{equation}
which is just the angular velocity of a simple mathematical
pendulum. This is even more evident from the second
time derivative:
\begin{equation}\label{dgl-pendulum1}
\ddot \w=2g\dot z=4vgy=4vgr\sin \w\,
\end{equation}
or with $\w=\pi+\uu$
\begin{equation}\label{dgl-pendulum2}
\ddot \uu+4vgr\sin \uu =0\,,
\end{equation}
which is the equation of motion for a mathematical pendulum, where
the ratio between gravitational acceleration and pendulum length is
replaced by $4vgr$, which is an energy dependent constant.
The two cases (i) and (ii) distinguished in section \ref{ss-viviani}
are simply the well-known librational and rotational motions
of the pendulum.\\[2mm]
(i) {\it Rotation:} With the abbreviations
\begin{equation}\label{dgl-pendulumb}
b=\frac{a^2+\R^2-1}{2a\R}\ ,\quad B=2g\sqrt{2a\R}\,.
\end{equation}
we obtain 
the period  as twice the integral over $\dot\theta^{-1}$
from $0$ to $\uu_0=\pi$  as
\begin{equation}\label{sol-periodR}
T_{\rm R}=\frac{1}{B}\int_0^{\pi}\frac{\rd \uu}{\sqrt{\cos \uu -b\,}}
=\frac{1}{g\sqrt{mar}}\,K(1/m)\,,
\end{equation}
(note that $|b|>1$) with 
\begin{equation}\label{parameter-m}
m=\frac{1-b}{2}=\frac{1-(\R-a)^2}{4a\R}\ge 1 
\end{equation}
and
$K$ is the complete elliptic integral of the first kind.\\[2mm] 
(ii) {\it Libration:} For $|b|\le 1$ we have a hindered rotation 
restricted to the interval $|\uu|\le \uu_0 =\arccos b$, 
and the period is $4$ times the integral over the interval $0<\uu<\uu_0$, that is, 
\begin{equation}\label{sol-periodL}
T_{\rm L}=\frac{4}{B}\int_0^{\uu_0}\frac{\rd \uu}{\sqrt{\cos \uu -b\,}}
=\frac{2}{g\sqrt{ar}}\,K(m)\,.
\end{equation}
where the parameter (\ref{parameter-m}) satisfies $|m|\le1$.

This, as well as the explicit construction of the solution in terms of Jacobi elliptic functions, 
is, of course, well known for the pendulum and also
for the Bose-Hubbard dimer \cite{Holt01a} 
(see, e.g., \cite{Reic90,Reic04}).
It should be noted, that also the integrals
of section \ref{s-area} appear as action integrals for the pendulum.
Some limiting cases may be of interest:\\[2mm]
(1)  For trajectories with $\R$ close to the boundary of
the allowed dynamical region, i.e.~for $\R \gtrsim a-1$  for $a>1$ or for
$\R \lesssim a+1$, we have $m \approx 0$ and with $K(0)=\pi/2$ 
the period (\ref{sol-periodL}) reduces to
\begin{equation}
T_{\rm L\pm}=\frac{\pi}{g\sqrt{a(a\mp 1)}}=
\frac{\pi}{\sqrt{v(v\mp g)}}\,,
\label{T_L}
\end{equation}
where one recognizes the celebrated Bogoliubov frequency 
$\Omega =2\pi/T=2\sqrt{v(v \mp  g)}$ for small  
excitations around the ground- and highest excited state, respectively  
\cite{Bogo47,Pita03}.

Clearly for this small angle oscillation this result can be directly
obtained from the pendulum equation (\ref{dgl-pendulum2}). On the Bloch 
sphere this is an oscillation in the vicinity of the fixed points
${\boldsymbol s}_{0\pm}=(\pm 1, 0,0)$ (compare eq.~(\ref{fixpoint})).\\[1mm]
(2) For $a<1$ and $\R=0$ we find the two fixed points 
${\boldsymbol s}_{1\pm}=(a,0,\pm\sqrt{1-a^2})$ from (\ref{fixpoint}). The oscillation
period in their vicinity can be found from (\ref{dgl-pendulum1}) as
\begin{equation}
T_{\rm R}=2g\sqrt{1-a^2}=2\sqrt{g^2-v^2}\,.
\label{T_R}
\end{equation}
(3) In the special case (b) mentioned in section
\ref{ss-viviani} we have $a<1$ and $\R=1-a$ (i.e.~$b=-1$ and $m=1$) the orbit approaches
the unstable fixed point ${\boldsymbol s}_{0+}=(1,0,0)$ along the separatrix and
the period becomes infinite.\\[2mm]
(4) For $\R=a$ the orbit passes through the poles (case (a) in section
\ref{ss-viviani}) and the period is given by
\begin{equation}\label{periodphoa}
T=\Bigg\{
\begin{array}{lll}
\frac{2}{g}\,K(4a^2) \quad &{\rm for}\quad a < \tfrac12 \\
\frac{2}{v}\,K(1/4a^2)\quad &{\rm for}\quad   a>\tfrac12
\end{array}
\end{equation}
as shown in figure \ref{fig_S_rho_a}.

We have shown that the restriction to the cylinder (\ref{eqn_cyl}) reduces 
the dynamics of the symmetric dimer  on the Bloch sphere to a simple pendulum 
motion on the level sets of constant energy. A similar construction
for Euler's equations for the free rigid body can be found in
\cite{Holm91,Holm08}.

\subsection{Full many-particle description}
\label{s-quantum}
In this section we will illustrate some implications of the
properties of the mean-field approximation discussed above for
the many-particle two-mode Bose-Hubbard Hamiltonian (\ref{BH-Hamiltonian})
with $\hbar=1$, where we confine ourselves to the
supercritical case $g>v$ ($v>0$, $g>0$).

\begin{figure}[b]
\begin{center}
\includegraphics[width=60mm]{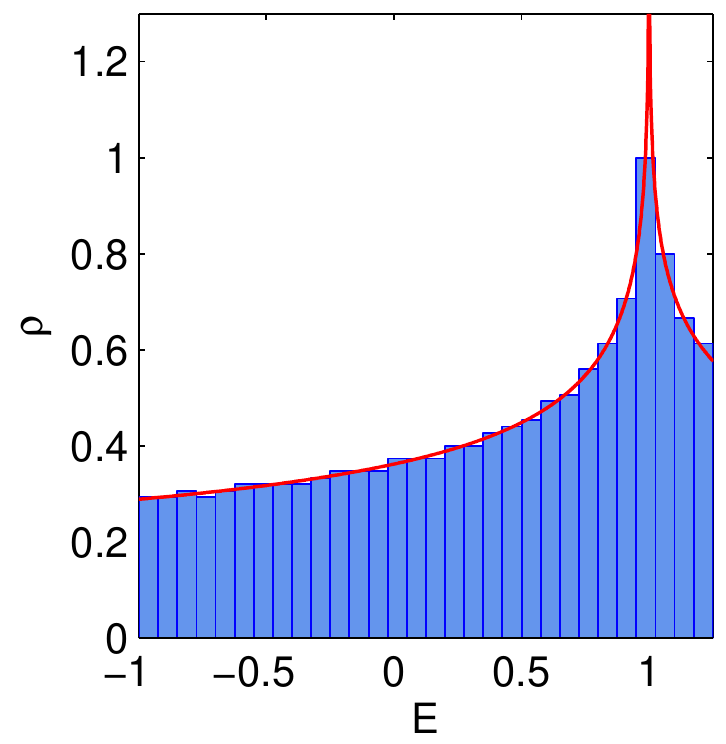}
\end{center}
\caption{\label{fig_density} Level density of the quantum
eigenvalues for $N=1000$ particles ($v=0.5$, $g=1$) in comparison with
the classical mean-field period $T$ (red line).} 
\end{figure}
\begin{figure}[t]
\begin{center}
\includegraphics[width=60mm]{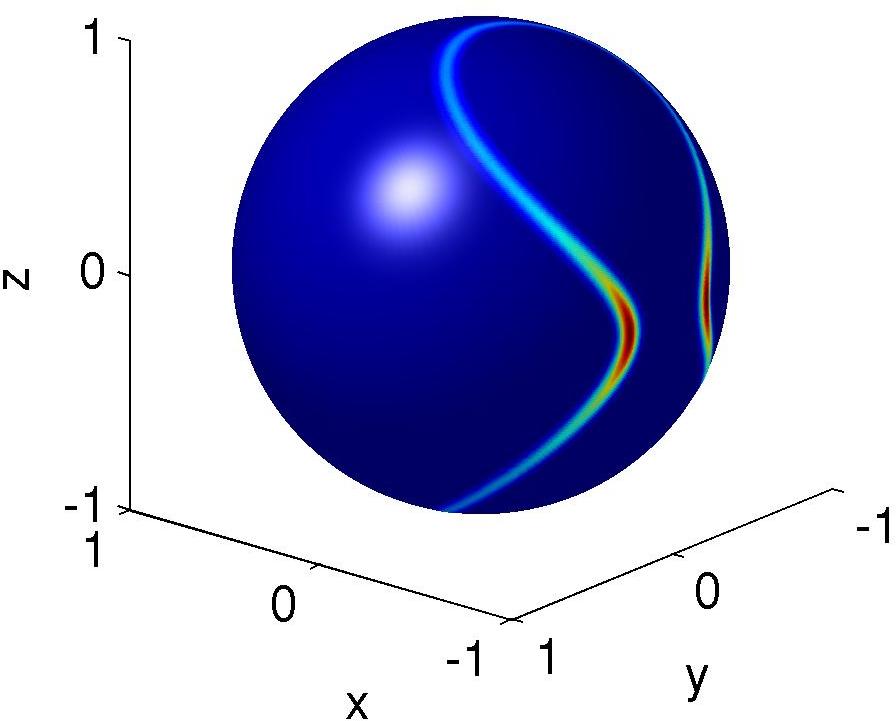}\hspace*{8mm}
\includegraphics[width=60mm]{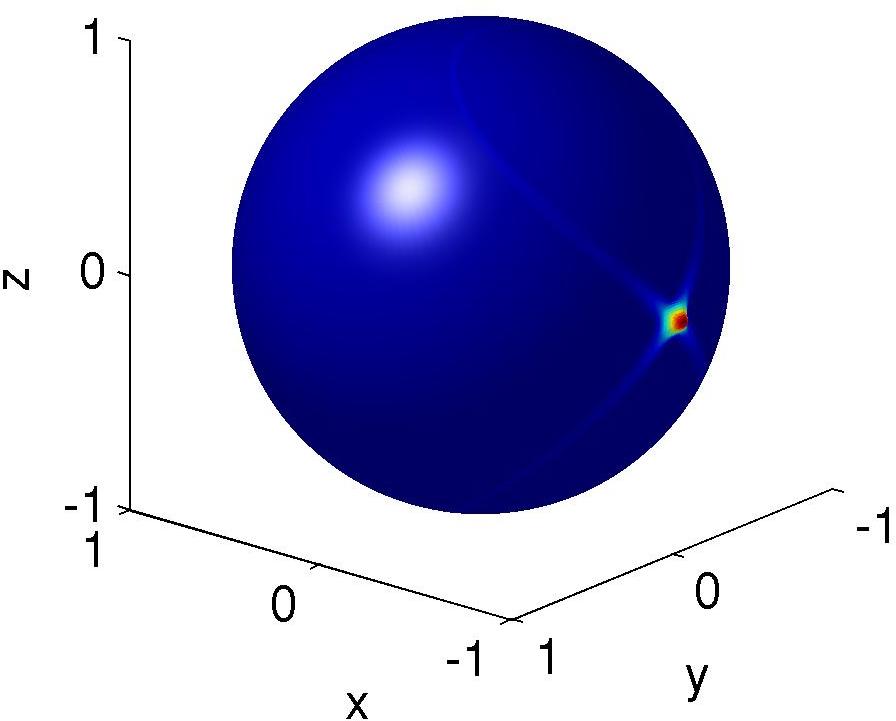}
\includegraphics[width=60mm]{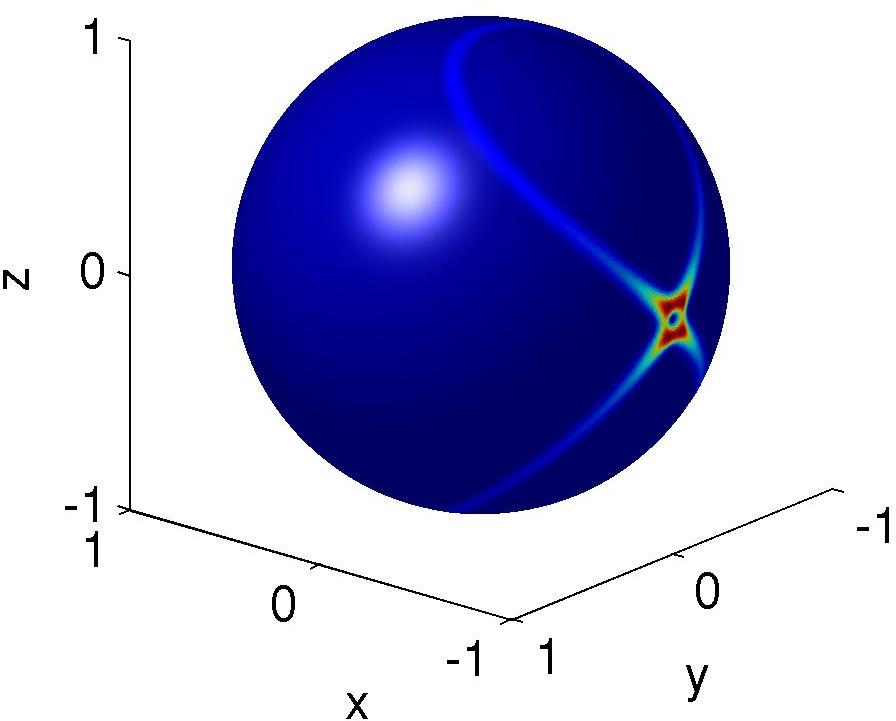}\hspace*{8mm}
\includegraphics[width=60mm]{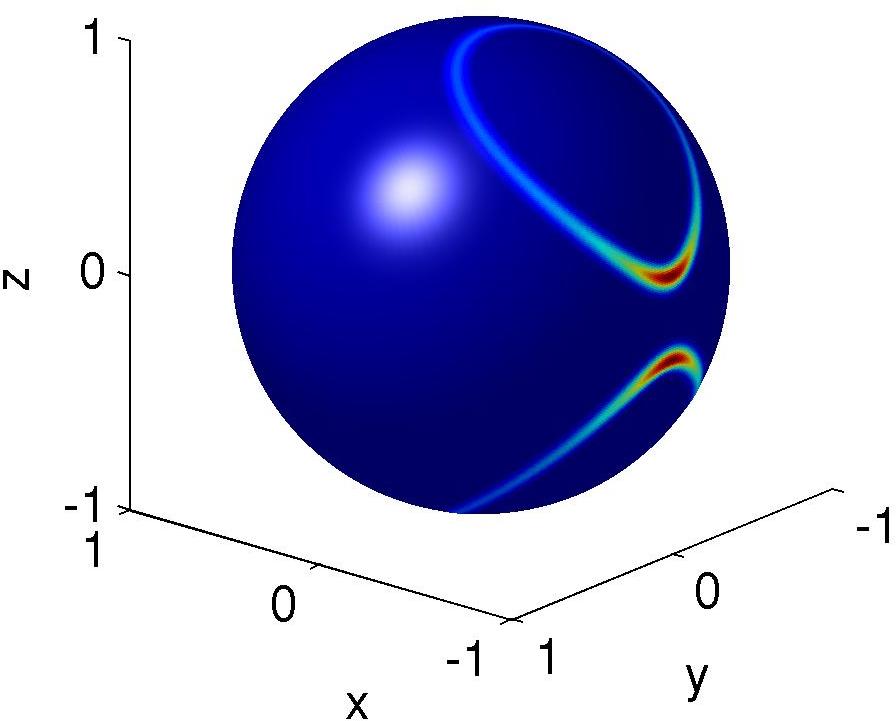}
\end{center}
\caption{\label{fig_husimi} Husimi distribution for eigenstates
$n=790$ (top left), $819$, $820$ and $837$
(bottom right) ($v=1$, $g=2$, $N=1000$).} 
\end{figure}

Diagonalizing the $N$-particle Hamiltonian we obtain the
$N+1$ energy eigenvalues $E_n$, $n=0,\ldots,N$ which are found
in the classically allowed mean-field interval
\begin{equation}
\label{E-interval}
-v<\frac{E_n}{N}<\frac{g}{2}\Big(1+\frac{v^2}{g^2}\Big)
\end{equation}
as discussed above.
Figure \ref{fig_density} shows the level density 
\begin{equation}
\rho=\frac{\Delta n}{\Delta E}
\end{equation}
as a function of the mean-field energy $E$
for $N=1000$ particles and $v=1$, $g=2$, where the energy interval
is discretized in 30 equidistant boxes $\Delta E$. 

Semiclassically the individual quantum energy eigenvalues $E_n$ can
approximately be calculated from the classical action integrals by
the Bohr-Sommerfeld quantization scheme (see, e.g., \cite{Gutz90}), 
i.e.~by the area $S(E)$ on the Bloch 
sphere enclosed by the classical orbit \cite{07semiMP,Simo12} and the 
level density is related to the energy derivative 
of the action, i.e.~the period $T$, see eq.~(\ref{T_L}) and (\ref{T_R}), 
which is also shown in the figure (compare also figure (\ref{fig_S_rho_a})).
The mean-field period $T(E)$ diverges for trajectories passing through
the saddle point, which agrees with the Viviani case for the
chosen parameter values.  For energies above the Viviani energy
$E_V=v=1$ the action area consists of two disconnected loops with
the same area and therefore the period is multiplied by a factor of two.
The Viviani action $S_V\approx 2.28$ in (\ref{S-viviani-1}) determines semiclassically
the number of states supported by the area of the Viviani window, i.e.~the
number $N _V\approx S_V(N+1)/(4\pi)\approx 182$ of states above the Viviani energy $E_V$.
Hence we expect $N+1-N_V=819$ states below the Viviani energy.

In the vicinity of this Viviani energy the quantum energy density shows
a pronounced maximum \cite{07semiMP,Chuc10}. For larger energies we have two almost degenerate
states with opposite symmetry. As an example figure \ref{fig_husimi}
shows the Husimi phase space distributions 
$|\langle \vartheta,\varphi|\Psi_n\rangle|^2$ of the
eigenstates $|\Psi_n\rangle$ on the Bloch sphere for $n=790$, $819$, $820$ and $837$
with energies 
$E(n)/N = 0.9733$, $1.0007$, $1.0017$, and $1.0158$. The first
state localizes on a single closed classical orbit,
the second one, $n=819$, almost exactly at the saddle point
in agreement with the semiclassical estimate above.
For the other two states the Husimi densities are localized on
the two disconnected loops encircling the
classical stationary points $\s_{1\pm}$ for the corresponding energies.

\begin{figure}[htb]
\begin{center}
\includegraphics[width=7cm]{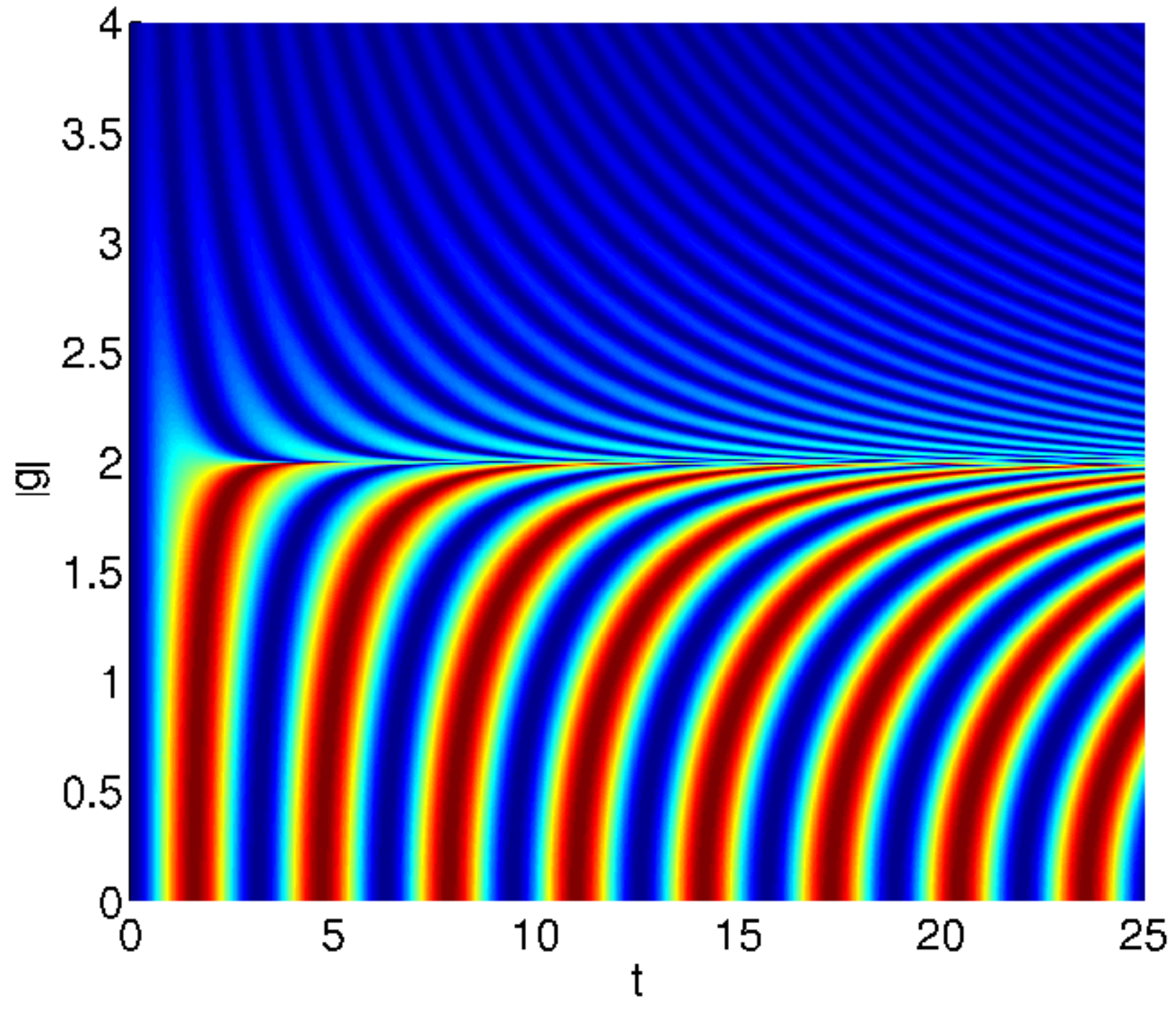}
\includegraphics[width=7.9cm]{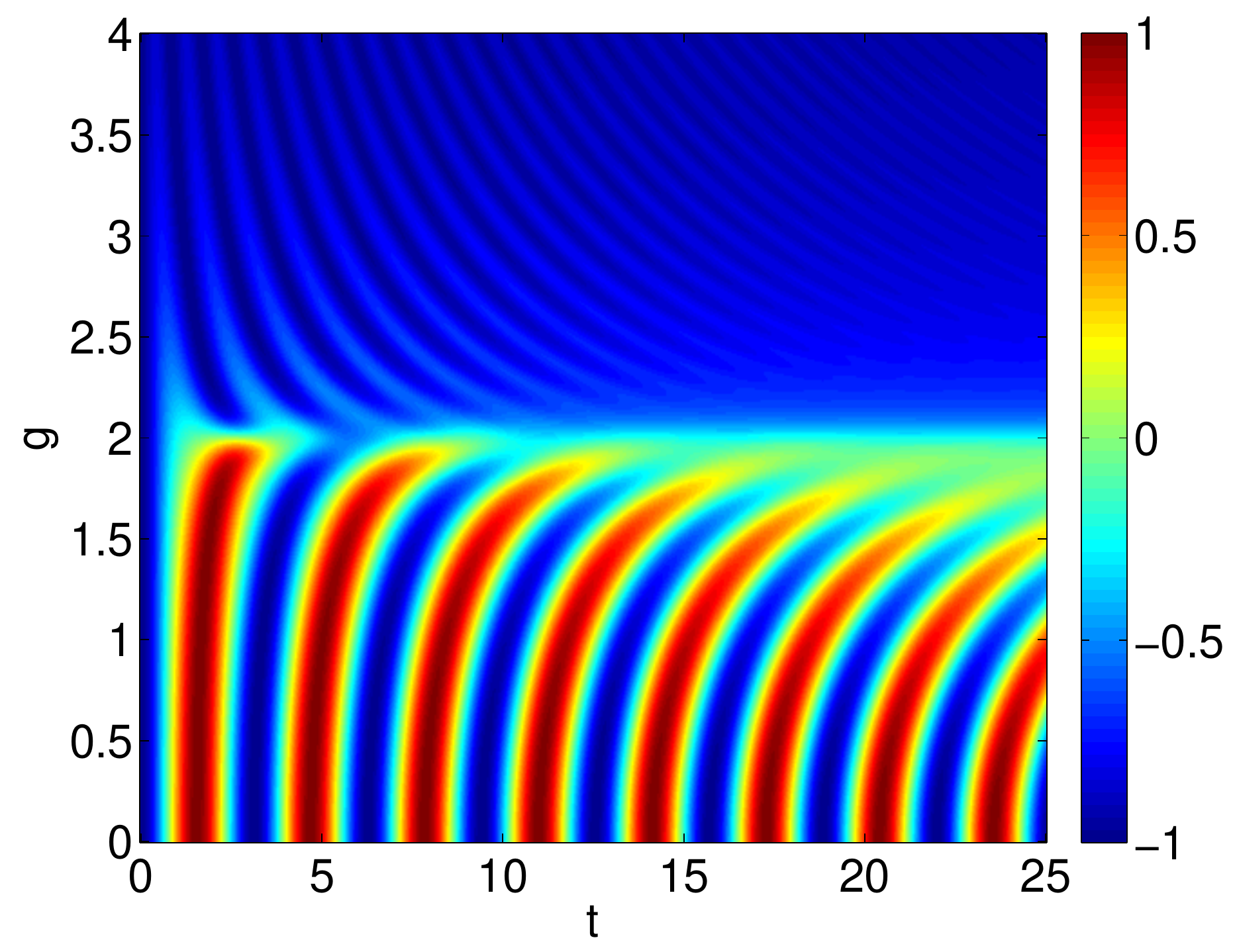}
\end{center}
\caption{\label{fig_MFST2}
Nonlinearity parameter dependence of the dynamics plotted in false colors  for 
initial conditions on the south pole of the Bloch sphere for $v=1$.
Left: Mean-field population imbalance $z=|\psi_1|^2-|\psi_2|^2$,
Right: Many particle expectation value of $\hat L_z$ for $N=1000$ particles.} 
\end{figure} 

Let us conclude with a brief discussion of the implications
of the classical Viviani curve
on the quantum dynamics for the important case
when the system is prepared at time $t=0$ in one of the modes. 
The left panel of figure \ref{fig_MFST2} shows 
the nonlinearity dependence of the time evolution of the mean-field population
imbalance $z=|\psi_1|^2-|\psi_2|^2$ (shown in false colors), where the system is
prepared in mode $2$ corresponding to the south pole of the Bloch
sphere \cite{10nhbh}. 
For vanishing nonlinearity one would recover the usual sinusoidal
Rabi oscillations between the two modes. For larger nonlinearities
the oscillation period increases but one still observes a
periodic complete population transfer between the modes.  
Although the self trapping bifurcation happens at $\g=v$
one only observes a characteristic change in the dynamics at 
the Viviani critical value of $\g=2v=2$. 
This can be understood in the following way: At
the self trapping transition a saddle point appears, but
the orbit passing through the north pole does not reach this point
and consequently does not 
change its
characteristics of a complete population transfer. With increasing 
nonlinearity, however, it coincides with the separatrix orbit
at the critical point $\g_{\rm}=2v$, the period of which
diverges, as can be nicely observed in figure \ref{fig_MFST2}. For
larger values we still find oscillatory
behavior, however, $z$ stays confined to negative values and the
system is therefore mainly populating the second mode and performs
self-trapping oscillations. 

Also shown in figure \ref{fig_MFST2} (right panel) is the 
corresponding quantum many-particle expectation value 
$2\langle\hat L_z\rangle/N$ for $N=1000$ particles
and an initial coherent state localized at the south pole. 
For vanishing interaction the mean-field description is exact
but for nonvanishing interactions one observes a decay of the
population imbalance,
in particular in the vicinity of the critical interaction.  
The mean-field
approximation is still restricted to the Bloch sphere whereas the
many-particle angular momentum expectation value can penetrate
the sphere. This breakdown of the mean-field approximation
\cite{Vard01b,Angl01} is a consequence of a mean-field representation
by a single phase space point and can be partly cured by  
the Liouville dynamics approach  
\cite{07phase,09phaseappl,Grae09dis,Henn12}, i.e. by averaging
over an ensemble of initial conditions
mimicking the Husimi distribution in classical phase space.
\section{Summary}
\label{sec_sum}
An interesting and unexpected interconnection between contemporary
cold atom quantum physics and much older studies in mathematics and
astronomy is provided by the celebrated Bose-Hubbard dimer.
We have identified the  mean-field trajectories of 
the symmetric Bose-Hubbard dimer as (generalized) Viviani curves or 
euclidic spherical ellipses. Furthermore we have shown that the
dynamics reduces to the oscillation of a mathematical pendulum
on a circle with an energy dependent radius as illustrated in figure
\ref{fig-circles12}.
 
In the librational case (ii) the pendulum motion
is restricted to an angular region inside the sphere in the $(x,y)$-plane
(the full line in the figure).
The corresponding motion in the $z$-direction extends from the 
northern to the southern hemisphere and the spherical ellipse is
a single closed loop.

For the rotational pendulum motion (case (i)) the spherical
ellipse consists of two loops on the northern and southern
hemisphere. The projection on the $z$-axis is restricted to an interval on the
positive or negative half-axis, respectively. In the language
of the Bose-Hubbard system, $z$ is the population imbalance and
therefore the population is trapped in a certain interval, an
effect known as self-trapping. The self-trapping transition
occurs for $g=v$, i.e.~when the radius $\R$ of the sphere
equals the shift $a$ of the center of the cylinder. 

The areas enclosed by the Viviani curves are thus the action integrals needed for a semiclassical 
quantization of the many-particle spectrum, and govern the energy density. The Viviani curve appears as dynamical separatrix between full oscillations and self-trapping oscillations when the system is prepared in one of the two modes.

\section*{Acknowledgments}
EMG gratefully acknowledges support via the Imperial College JRF scheme. 

\section*{References}
\bibliographystyle{unsrtot}
\bibliography{abbrev,publko,paper60,paper70,paper80,paper90,paper00,rest,dipldiss}

\begin{thebibliography}{10}

\bibitem{Milb97}
G.~J. Milburn, J.~Corney, E.~M. Wright, and D.~F. Walls,  Phys. Rev. A  {\bf
  55}  (1997)   4318

\bibitem{Smer97}
A.~Smerzi, S.~Fantoni, S.~Giovanazzi, and S.~R. Shenoy,  Phys. Rev. Lett.  {\bf
  79}  (1997)   4950

\bibitem{Ragh99}
S.~Raghavan, A.~Smerzi, S.~Fantoni, and S.~R. Shenoy,  Phys. Rev. A  {\bf 59}
  (1999)   620

\bibitem{Vard01b}
A.~Vardi and J.~R. Anglin,  Phys. Rev. Lett.  {\bf 86}  (2001)   568

\bibitem{Angl01}
J.~R. Anglin and A.~Vardi,  Phys. Rev. A  {\bf 64}  (2001)   013605

\bibitem{Link06}
J.~Links, A.~Foerster, A.~P. Tonel, and G.~Santos,  Ann. Henri Poincar\'e  {\bf
  7}  (2006)   1591

\bibitem{07semiMP}
E.~M. Graefe and H.~J. Korsch,  Phys. Rev. A  {\bf 76}  (2007)   032116

\bibitem{Grae09dis}
E.-M. Graefe,  {\it Quantum-Classical Correspondance for a Bose-Hubbard dimer
  and its non-Hermitian generalisation},, Dissertation, Techn. Univers.
  Kaiserslautern, Germany, 2009

\bibitem{Bouk09}
E.~Boukobza, M.~Chuchem, D.~Cohen, and A.~Vardi,  Phys. Rev. Lett.  {\bf 102}
  (2009)   180403

\bibitem{Chuc10}
M.~Chuchem, K.~Smith-Mannschott, M.~Hiller, T.~Kottos, A.~Vardi, and D.~Cohen,
  Phys. Rev. A  {\bf 82}  (2010)   053617

\bibitem{Niss10}
F.~Nissen and J.~Keeling,  Phys. Rev. A  {\bf 81}  (2010)   063628

\bibitem{Simo12}
L.~Simon and W.~T. Strunz,  Phys. Rev. A  {\bf 86}  (2012)   053625

\bibitem{Viviani}
see, e.g., \url{http://en.wikipedia.org/wiki/Viviani%27s_curve} or
  \url{http://mathworld.wolfram.com/VivianisCurve.html}

\bibitem{Hippo}
see, e.g., \url{https://en.wikipedia.org/wiki/Eudoxus_of_Cnidus} or \url{
  http://demonstrations.wolfram.com/HippopedeOfEudoxus/}

\bibitem{cadd01}
R.~Caddeo, S.~Montaldo, and P.~Piu,  Math. Intelligencer  {\bf 23(3)}  (2001)
  36

\bibitem{Krol05}
W.~Kroll,  Der Mathematikunterricht  {\bf 51(6)}  (2005)   15

\bibitem{Krol07}
W.~Kroll,  {\em R\"aumliche Kurven und Fl\"achen in ph\"anomenologischer
  Behandlung},   (self-published, ISBN 978-3-00-021836-1), 2007

\bibitem{Yave01}
I.~Yavetz,  Arch. Hist. Exact Sci.  {\bf 56}  (2001)   69

\bibitem{Albi05}
M.~Albiez, R.~Gati, J.~F\"olling, S.~Hunsmann, M.~Cristiani, and M.~K.
  Oberthaler,  Phys. Rev. Lett.  {\bf 95}  (2005)   010402

\bibitem{Zibo10}
T.~Zibold, E.~Nicklas, C.~Gross, and M.~K. Oberthaler,  Phys. Rev. Lett.  {\bf
  105}  (2010)   204101

\bibitem{Lipk65}
H.~J. Lipkin, N.~Meshkv, and A.~J. Glick,  Nuclear Phys.  {\bf 62}  (1965)
  188

\bibitem{Mesh65}
N.~Meshkov, A.~J. Glick, and H.~J. Lipkin,  Nuclear Phys.  {\bf 62}  (1965)
  199

\bibitem{Glic65}
A.~J. Glick, H.~J. Lipkin, and N.~Meshkv,  Nuclear Phys.  {\bf 62}  (1965)
  211

\bibitem{Vida04b}
J.~Vidal, G.~Palacois, and C.~Aslangul,  Phys. Rev. A  {\bf 70}  (2004)
  062304

\bibitem{Lato05}
J.~I. Latorre, R.~Or\'us, E.~Rico, and J.~Vidal,  Phys. Rev. A  {\bf 71}
  (2005)   064101

\bibitem{Ribe07}
P.~Ribeiro, J.~Vidal, and R.~Mosseri,  Phys. Rev. Lett.  {\bf 99}  (2007)
  050402

\bibitem{Orus08}
R.~Or\'us, S.~Dusuel, and J.~Vidal,  Phys. Rev. Lett.  {\bf 101}  (2008)
  025701

\bibitem{Dira27}
P.~A.~M. Dirac,  Proc. Roy. Soc. Lond. A  {\bf 114}  (1927)   243

\bibitem{Wein89}
S.~Weinberg,  Ann. Phys. (N.Y.)  {\bf 194}  (1989)   336

\bibitem{Kenk86}
V.~M. {Kenkre and D. K. Campbell},  Phys. Rev. B  {\bf 34}  (1986)   4959

\bibitem{Holt01a}
M.~Holthaus and S.~Stenholm,  Eur. Phys. J. B  {\bf 20}  (2001)   451

\bibitem{Reic90}
L.~E. Reichl and Li~Haoming,  Phys. Rev. A  {\bf 42}  (1990)   4543

\bibitem{Reic04}
L.~E. Reichl,  {\em The Transition to Chaos},   Springer, New York, 2004

\bibitem{Bogo47}
N.~N. Bogoliubov,  J. Phys. USSR  {\bf 11}  (1947)   23

\bibitem{Pita03}
L.~Pitaevskii and S.~Stringari,  {\em Bose-Einstein Condensation},   Oxford
  University Press, Oxford, 2003

\bibitem{Holm91}
D.~D. Holm and J.~E. Marsden,  in {\em Symplectic Geometry and Mathematical
  Physics}, edited by P.~Donato, C.~Duval, J.~Elhadad, and G.~M. Tuynman.
  Birkh\"auser, Boston, 1991

\bibitem{Holm08}
D.~D. Holm,  {\em Geometric Mechanics Part I: Dynamics and Symmetry},
  Imperial College Press, London, 2008

\bibitem{Gutz90}
M.~C. Gutzwiller,  {\em Chaos in Classical and Quantum Mechanics},   Springer,
  New York, 1990

\bibitem{10nhbh}
E.-M. Graefe, H.~J. Korsch, and A.~Niederle,  Phys. Rev. A  {\bf 82}  (2010)
  013629

\bibitem{07phase}
F.~Trimborn, D.~Witthaut, and H.~J. Korsch,  Phys. Rev. A  {\bf 77}  (2008)
  043631

\bibitem{09phaseappl}
F.~Trimborn, D.~Witthaut, and H.~J. Korsch,  Phys. Rev. A  {\bf 79}  (2009)
  013608

\bibitem{Henn12}
H.~Hennig, D.~Witthaut, and D.~K. Campbell,  Phys. Rev. A  {\bf 86}  (2012)
  051604

\end{thebibliography}
\end{document}